\PassOptionsToPackage{table}{xcolor}
\PassOptionsToPackage{dvipsnames}{xcolor} 
\documentclass[sigconf,natbib=true]{acmart}

\AtBeginDocument{%
  }

\usepackage{tcolorbox}

\usepackage{bbding}
\usepackage{subfigure}
\usepackage[boxed,ruled,lined]{algorithm2e}
\usepackage{makecell}

\newcommand{\argmin}{\operatornamewithlimits{arg\,min}}
\usepackage{arydshln} 
\usepackage{enumerate}
\usepackage{enumitem}
\usepackage{makecell}
\usepackage{bbm}
\usepackage{bm}
\newtheorem{assumption}{Assumption}

\usepackage[table]{xcolor}
\usepackage{booktabs}
\usepackage{multirow}

\definecolor{warmc}{HTML}{B8645C}
\definecolor{coldc}{HTML}{627CB9}
\definecolor{coldbg}{RGB}{232,242,255}   
\definecolor{coldtxt}{RGB}{0,82,155} 
\newcommand{\coldcell}[1]{\textcolor{coldtxt}{#1}}

\newcommand{\ourname}{GenRecEdit\xspace}

\setcopyright{acmlicensed}

\copyrightyear{2026}
\acmYear{2026}
\setcopyright{cc}
\setcctype{by}
\acmConference[SIGIR '26]{Proceedings of the 49th International ACM SIGIR Conference on Research and Development in Information Retrieval}{July 20--24, 2026}{Melbourne, VIC, Australia}
\acmBooktitle{Proceedings of the 49th International ACM SIGIR Conference on Research and Development in Information Retrieval (SIGIR '26), July 20--24, 2026, Melbourne, VIC, Australia}
\acmDOI{10.1145/3805712.3809662}
\acmISBN{979-8-4007-2599-9/2026/07}

\begin{document}

\title{\ourname: Adapting Model Editing for Generative
Recommendation with Cold-Start Items }

\author{Chenglei Shen}
\affiliation{
  \institution{Gaoling School of Artificial Intelligence, Renmin University of China}
  \city{Beijing}
  \country{China}
}
\email{chengleishen9@ruc.edu.cn}

\author{Teng Shi}
\affiliation{
  \institution{Gaoling School of Artificial Intelligence, Renmin University of China}
  \city{Beijing}
  \country{China}
}
\email{shiteng@ruc.edu.cn}

\author{Weijie Yu}
\affiliation{
  \institution{School of Information Technology and Management, University of International Business and Economics}
  \city{Beijing}
  \country{China}
}
\email{yu@uibe.edu.cn}

\author{Xiao Zhang}
\authornote{Corresponding author.}
\affiliation{
  \institution{Gaoling School of Artificial Intelligence, Renmin University of China}
  \city{Beijing}
  \country{China}
}
\email{zhangx89@ruc.edu.cn}

\author{Jun Xu}
\affiliation{
  \institution{Gaoling School of Artificial Intelligence, Renmin University of China}
  \city{Beijing}
  \country{China}
}
\email{junxu@ruc.edu.cn}

\renewcommand{\shortauthors}{Teng Shi et al.}

 
\begin{abstract}
Generative recommendation (GR) has demonstrated substantial potential for sequential recommendation in an end-to-end generation paradigm. However, existing models suffer from severe cold-start collapse, i.e., recommendation accuracy on cold-start items drops to near zero. Current solutions rely on retraining with cold-start interactions, which faces sparse feedback, high computational cost, and delayed updates, diminishing practical utility in rapidly evolving catalogs in recommendation. Inspired by NLP model editing (enabling training-free knowledge injection into large language models), we explore applying this paradigm to generative recommendation, but face two fundamental challenges: (1)~sequential data in GR lacks explicit subject–object binding (a core NLP sentence structure), hindering targeted model edits; (2)~sequential data in GR has no fixed token co-occurrence patterns (unlike NLP’s stable phrases), making multi-token injection unreliable. To address these, we propose GenRecEdit, the first model editing framework tailored for generative recommendation. Specifically, we: (1)~mitigate the absence of sentence structure by explicitly modeling the intrinsic relationship between the entire sequence context and the next-token (e.g., semantic IDs, SIDs) generation; (2)~adopt iterative token-level edits to effectively inject token bundles (e.g., items); and (3)~introduce a One-One trigger mechanism to avoid interactions among multiple token-level edits during inference. Extensive experiments across multiple datasets demonstrate that \ourname substantially improves recommendation performance on cold-start items while preserving the model’s original recommendation quality. Moreover, \ourname achieves these gains with only approximately \textbf{9.5\%} of the training time required by retraining, significantly reducing computational cost and enabling efficient, frequent model updates.

\end{abstract}




\maketitle

\section{Introduction}
\label{sec:intro}

\begin{figure}[h]
\centering
\includegraphics[width=0.98\columnwidth]{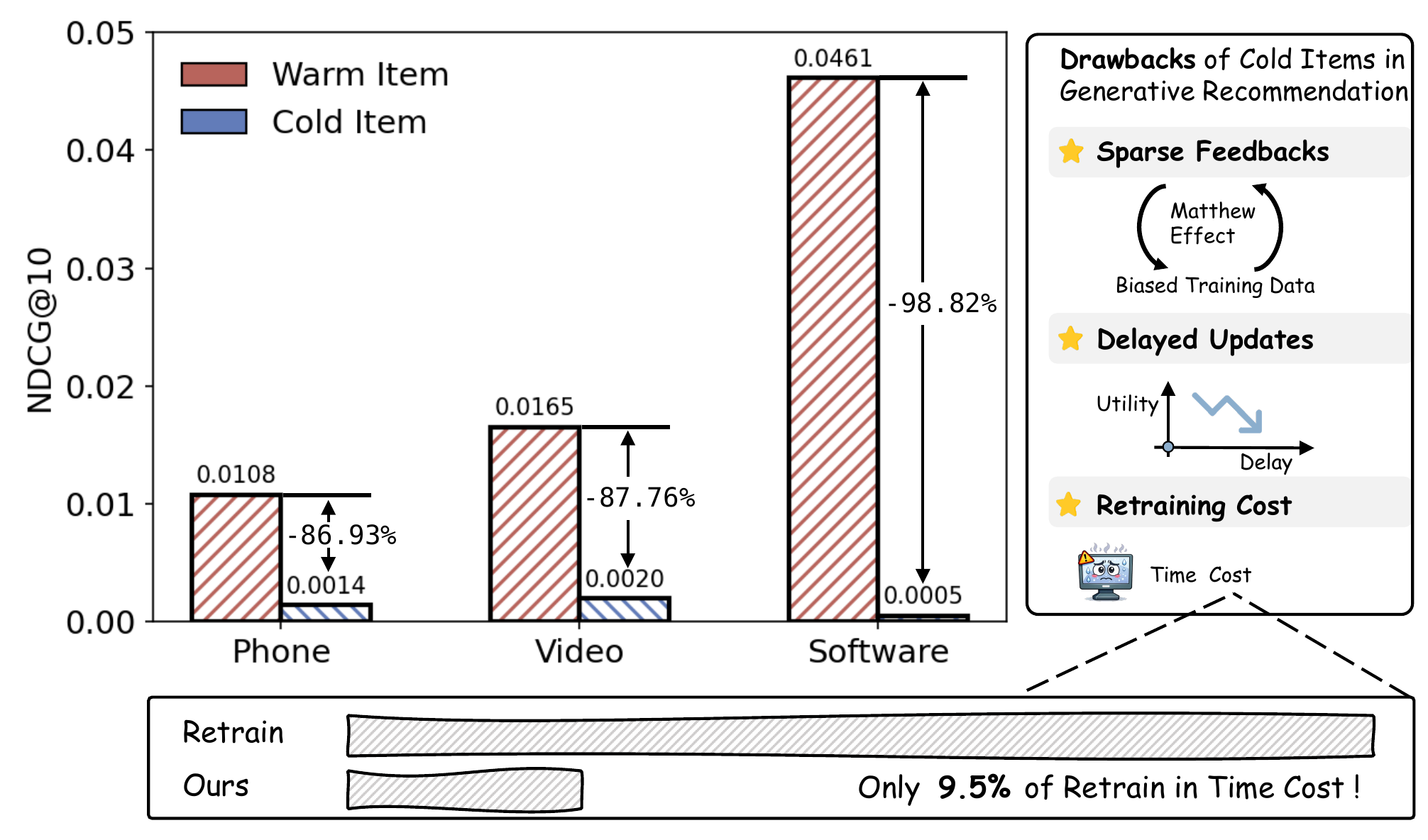}
\caption{
An illustration of cold-start collapse in GR. The left panel presents dataset statistics that characterize cold-start collapse, while the right panel summarizes the adverse effects on cold-start items. The bottom panel shows the time-efficiency of \ourname in terms of model update cost.
}
\label{fig:intro_necessary}
\vspace{-0.3cm}
\end{figure}

Generative recommendation (GR) has recently emerged as a promising paradigm for sequential recommendation tasks~\cite{TIGER,LETTER,LC_Rec,shi2025llada}. In GR, each item is tokenized into a small number of discrete semantic tokens (semantic IDs), and a sequential model is trained to autoregressively generate the next tokens, which are then parsed to form the predicted item. By reformulating recommendation as a sequence-to-sequence generation problem, GR departs from conventional discriminative sequential recommendation methods~\cite{SASREC,sun2019bert4rec,qu2025bridging}, enabling improved scalability and superior performance~\cite{zhou2025onerec,deng2025onerec}, while directly leveraging the optimization advances of large language models~\cite{zhai2024actions,qin2025maps,qin2025more,shen2026enhancing}.

Unlike item-ID–based transductive models~\cite{SASREC,sun2019bert4rec,shen2023hyperbandit}, which are unable to recommend newly introduced items due to the absence of their IDs in the trained model, generative recommendation (GR) models are designed to generalize beyond the closed item vocabulary by exploiting semantic correlations and generating semantic tokens for previously unseen items (i.e., cold-start items). However, in practice, GR models often perform poorly on cold-start items. As illustrated in Figure~\ref{fig:intro_necessary}, we partition the test set into warm and cold subsets based on whether each test item appears in the training data. Evaluation with well-trained GR models reveals a striking failure mode: when cold-start items are introduced after training, recommendation accuracy on cold-start items can drop to nearly zero, a phenomenon we term \textbf{cold-start collapse} in GR.

To investigate this issue, we first pose a fundamental question: \textbf{What causes this collapse?} Our analysis in Section~\ref{sec:analysis} leads to two key findings. (1)~Well-trained GR models can often correctly generate the first semantic ID token of cold-start items, indicating substantial latent potential for cold-start recommendation. (2)~Regardless of whether the final recommendation is correct, GR models exhibit a strong bias toward generating seen semantic ID patterns~\cite{yang2024unifying,ding2026inductive}. A straightforward solution is to collect feedback on cold-start items and then retrain or incrementally finetune the model. However, this paradigm is limited in practice. For example, in scenarios where up-to-date recommendations are critical (e.g., news or short-video platforms), incremental training or retraining faces substantial challenges, including sparse feedback for cold-start items, delayed model updates~\citep{hou2025towards}, and prohibitive retraining costs. These findings and constraints underscore an urgent need for a method that enables on-the-fly incorporation of semantic ID patterns for cold-start items without compromising the inherent recommendation capability of GR models.

\begin{figure}[t]
\centering
\includegraphics[width=1.0\columnwidth]{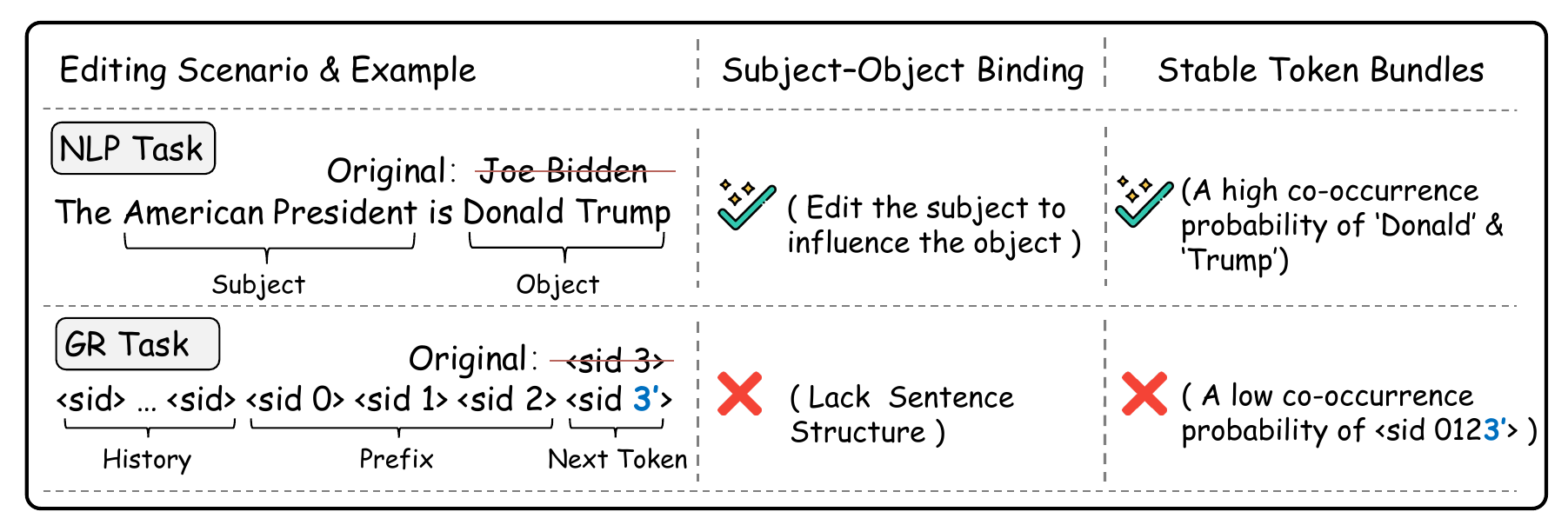}
\vspace{-5px}
\caption{
An illustration of the challenges adapting model editing from NLP
to generative recommendation (GR). 
}
\label{fig:example}
\end{figure}

Based on the above analysis, we naturally ask: \textbf{How to efficiently resolve this collapse?} Motivated by recent progress in model editing for natural language processing, which enables training-free knowledge injection into large language models, we explore whether a similar paradigm can be applied to generative recommendation by treating the semantic ID patterns of cold-start items as editable knowledge and injecting them into GR models. However, directly transferring existing model editing techniques to sequential recommendation data faces two fundamental challenges, as shown in Figure~\ref{fig:example}. \textbf{First}, GR sequences lack explicit sentence structure, especially clear subject--object binding, which makes it difficult to control the target item by editing a well-defined subject representation. For example, in NLP task, if we aim to change the American president from ``Joe Biden'' to ``Donald Trump'' given the context ``The American president is~\ldots'', a common strategy is to localize the subject phrase (``American president'') via the subject--object binding, and edit its hidden representation so that decoding favors the desired object tokens (``Donald Trump''). In GR, however, such a clear subject--object binding is typically absent due to the lack of explicit sentence structure.  \textbf{Second}, natural language often contains stable token bundles (i.e., phrases) with high co-occurrence probability (e.g., `Donald' and `Trump'), benefiting from extensive corpus-level pretraining. In contrast, GR sequences do not exhibit fixed co-occurrence patterns for cold-start items because the model has not observed their SID patterns during training. Consequently, injecting multi-token sequences becomes unreliable.

Therefore, the key question is: \textbf{How to overcome these challenges?} To this end, we propose \ourname, a model editing framework tailored for generative recommendation. Specifically, \textbf{(1)} we mitigate the absence of sentence structure by explicitly modeling the intrinsic relationship between the entire sequence context and the next-token (e.g., a single semantic ID); \textbf{(2)} we adopt iterative token-level edits (i.e., position-wise edits) to effectively inject token bundles (i.e., semantic ID patterns of cold-start items); and \textbf{(3)} we introduce a One-One trigger mechanism to avoid interactions among multiple token-level edits during inference.

We summarize our contributions as follows:
\begin{itemize}[leftmargin=*]
    \item We reveal a striking cold-start collapse in generative recommendation: accuracy on cold-start items can drop to near zero, and our analysis shows that GR models have the potential to recommend cold-start items but default to seen semantic-ID patterns (Section~\ref{sec:analysis}, Figure~\ref{fig:intro_necessary}).
    
    \item We propose \ourname, which treats cold-start semantic-ID patterns as editable knowledge and injects them into GR models in a training-free, on-the-fly manner, while preserving the model’s existing recommendation capability.
    
    \item Extensive experiments on multiple datasets demonstrate that \ourname substantially improves cold-start recommendation with minimal impact on the original general recommendation ability, while incurring only \textbf{9.5\% }of the retraining time cost.

\end{itemize}

\section{Related Work}

\noindent\textbf{Generative Recommendation.} Motivated by the strong empirical performance of large language models~(LLMs)~\cite{zhao2023survey}, generative recommendation has attracted growing interest~\cite{hua2023index,TIGER,LC_Rec,LETTER,RPG,deng2025onerec,zhou2025onerec}. In this paradigm, each item is represented as a sequence of discrete semantic IDs (SIDs); the interaction history is then a concatenation of item SIDs, and a generative model predicts the target item’s SIDs. Generative recommendation typically decouples into item tokenization and autoregressive generation: tokenization assigns SIDs from item meta semantics, while the autoregressive model learns intra-item SID patterns and inter-item sequential dependencies.  
Recent advances suggest model intervention improves the test-time controllability of recommendation models \citep{shen2026survey}. Prior work has studied test-time model generation \citep{chen2023controllable,shen2025paragon} and test-time training \citep{yang2024ttt4rec,zhang2025test}, yet test-time intervention for generative recommendation with cold-start items is still underexplored. 
Existing tokenization schemes (e.g., clustering~\cite{si2024generative,wang2024eager} and vector quantization~\cite{TIGER,deng2025onerec,zhou2025onerec,zhu2024cost}) allow SIDs for cold-start items to be constructed offline, alleviating the embedding-missing issue in ID-based recommenders. However, as highlighted in prior studies~\citep{hou2025towards,ding2026inductive}, a key bottleneck remains: generative recommendation models predominantly generate items seen during training and exhibit limited ability to recommend unseen items, primarily because the \emph{SID patterns} of cold-start items are absent from the training data. Similarly, inspired by the future research directions outlined in the generative recommendation tutorial~\citep{hou2025towards}, we address this by applying model editing to inject cold-start SID patterns while preserving previously learned knowledge, thereby improving cold-start generation accuracy.
\\

\noindent\textbf{Model Editing.} Model editing has emerged as a lightweight, training-free paradigm for inserting, modifying, and removing knowledge in large language models~\citep{meng2022mass,fang2024alphaedit}. It is motivated by the view that knowledge is propagated by attention and stored in linear layers, especially feed-forward networks (FFNs), which are thus the main targets for intervention~\citep{geva2021transformer,meng2022mass,meng2022locating}. Most methods follow a locate-then-edit pipeline: they first localize the edit layer and then exploit the linear mapping from FFN hidden-state shifts to weight updates. Representative approaches include ROME~\citep{meng2022locating}, which analyzes how factual knowledge is encoded and shows that internal mechanisms can be directly manipulated; MEMIT~\citep{meng2022mass}, which enables large-scale edits at specific layers while maintaining specificity and fluency; and $\alpha$-Edit~\citep{fang2024alphaedit}, which mitigates the update--preservation trade-off via null-space projection. AnyEdit~\citep{jiang2025anyedit} extends editing to unstructured knowledge through token-level hidden-state edits, but remains focused on textual, non-lifelong settings.  
Notably, several recent studies have further substantiated the validity of this technical paradigm in information retrieval. As a pioneering work that first extends model editing to generative retrieval, DOME~\citep{zhang2026model} tackles the critical issue of indistinguishable edit vectors for new document integration, and learns discriminative edit vectors to enable efficient, low-forgetting adaptation of generative retrieval models. 
RAIE~\citep{zeng2026raie} introduces knowledge editing-inspired adaptation for LLM-based recommendation, focusing on region-aware incremental preference updates to tackle user preference drift. 
With the rise of generative recommendation, the boundary between recommender systems and large language models is increasingly blurred, and cold-start items naturally resemble newly introduced knowledge in LLMs. To address this challenge, we propose \ourname, an effective approach to apply model editing to generative recommendation.

\section{Problem Analysis and Formulation}

This section analyzes the causes of cold-start collapse in Generative Recommendation (GR) and formulates the problem of model editing for GR.
\subsection{Problem Analysis}
\label{sec:analysis}
\begin{figure}[t]
\centering
\includegraphics[width=1.0\columnwidth]{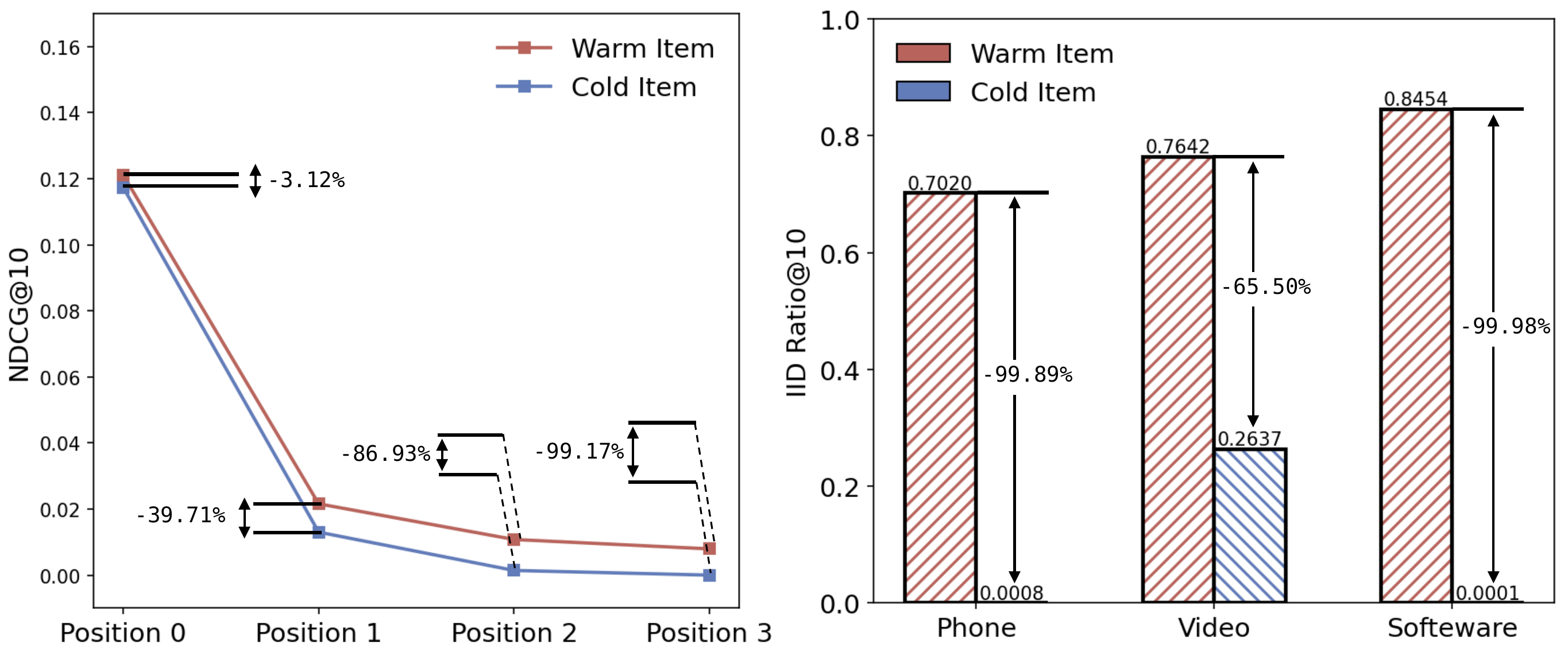}
\caption{
The analysis of the cold-start collapse. \textbf{Left:} NDCG at the first $n$ positions when a GR model generates a four-position semantic ID on the  \emph{Cell Phones and Accessories} dataset. \textbf{Right:} Distribution of generated items regardless of recommendation correctness; a higher IID Ratio indicates a larger fraction of generated items that belong to the current test subset (i.e., cold subset or warm subset).
}
\label{fig:intro_example}
\end{figure}


We focus on a key question: \textbf{what causes the cold-start collapse in GR?} Our analysis suggests that GR models retain substantial potential for recommending cold-start items. The primary bottleneck is not a lack of general recommendation capability, but rather the model’s inability to generate semantic ID patterns that it has rarely observed for cold-start items. To substantiate this claim, we design two experiments. (1) Instead of treating a full four-token semantic ID as an item, we treat the prefix consisting of the first $n$ semantic-ID tokens as an item and examine how NDCG changes when $n \in \{0,1,2,3\}$. The results are shown in Figure~\ref{fig:intro_example} (Left). (2) We define a metric termed $\mathrm{IID~Ratio@K}$, which ignores recommendation correctness and measures the fraction of the top-$K$ generated items that belong to in-distribution items (e.g., the whole
cold-start items for cold subset) in the current test set. Formally,
\begin{equation}
\mathrm{IID\ Ratio@K} = \frac{1}{|\mathcal{U}|} \sum_{u \in \mathcal{U}}
\frac{\left| R_{iid} \cap \hat{R}_{u,K} \right|}{\left| R_{iid} \right|},
\end{equation}
where $\mathcal{U}$ denotes the set of users, $R_{iid}$ is the item set associated with the current test split, and $\hat{R}_{u,K}$ is the set of top-$K$ items generated by the GR model for user $u$. The results are shown in Figure~\ref{fig:intro_example} (Right).

These two experiments lead to the following observations. (1)~The first-token recommendation accuracy on cold-start items is relatively high, indicating that the GR model can often identify the coarse category of a cold-start item and generate a plausible initial semantic-ID token. However, the generation becomes increasingly unstable as the model proceeds to complete the full semantic ID. (2)~$\mathrm{IID Ratio@K}$ reveals the source of this instability: the large gap between warm and cold-start settings indicates that the model rarely generates cold-start items at all, regardless of whether the generated recommendation is correct.

\begin{figure*}[t]
\centering
\includegraphics[width=1.0\textwidth]{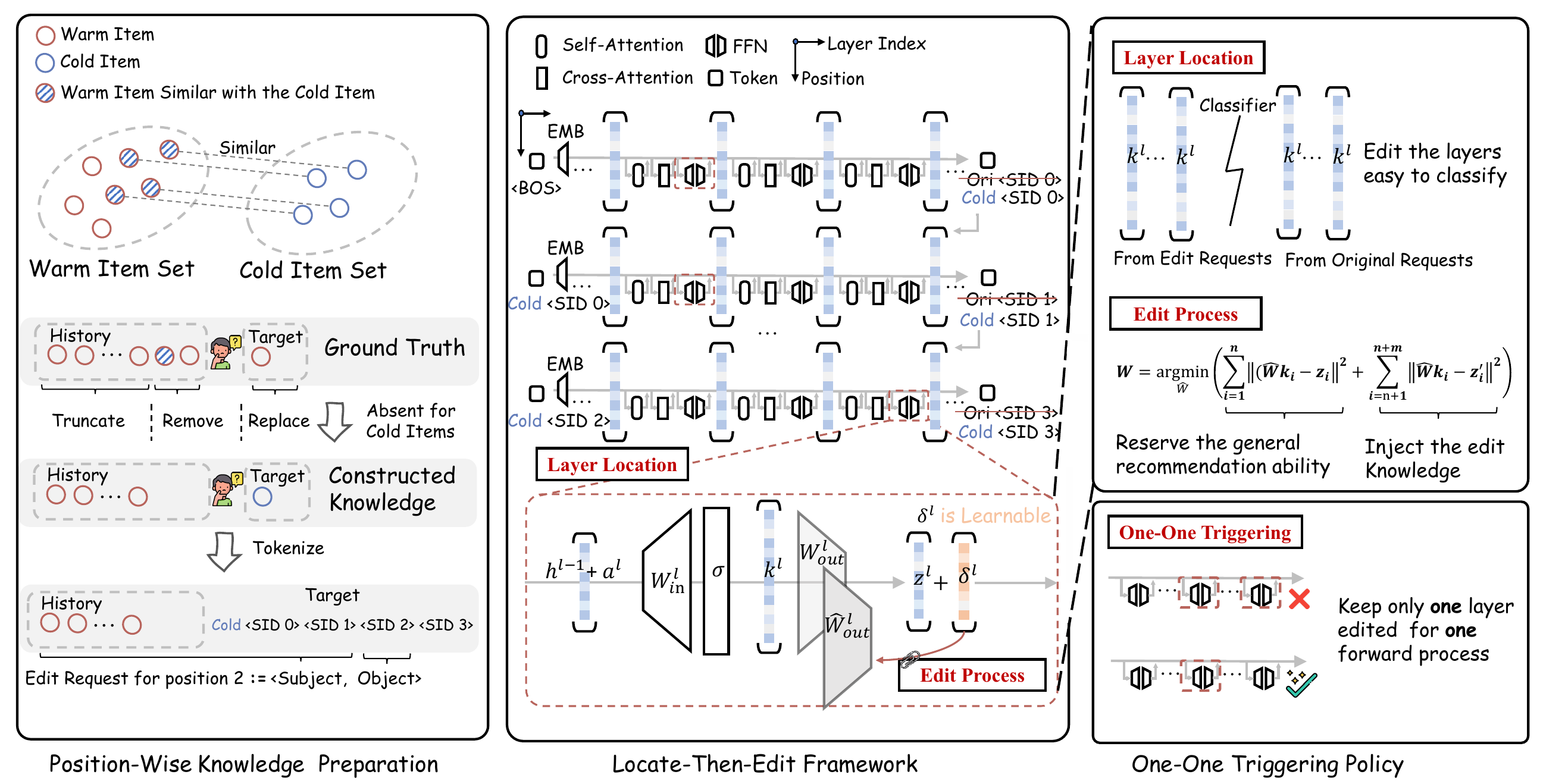}
\vspace{-8px}
\caption{ Overall framework of \ourname, which consists of three main modules: (1) Position-Wise Knowledge Preparation. We construct pseudo interaction data for cold-start items and use it to form edit requests. (2) Locate-Then-Edit Framework. For each position, we first use a classifier to localize the key layer that is most strongly associated with that position, and then perform targeted edits on the identified layer. (3) One-One Triggering Policy. To prevent interference among edits for different positions, we adopt an adaptive triggering strategy at inference time, where the corresponding edit layer is triggered according to the current decoding position.}
\label{fig:method}
\vspace{-0.3cm}
\end{figure*}

\subsection{Adapting Model Editing to GR}
\label{sec:GRmodelediting}
Model editing aims to update a model’s behavior on a small set of specified inputs without retraining. In a standard formulation, the edit requests contain a set of input (i.e., $context$) together with desired target outputs, and the goal is to modify model parameters so that the model produces the new targets while preserving its behavior elsewhere. Formally, an edit request is typically specified as a subject, relation, and object triple, denoted as $\langle s,r,o\rangle$. Our objective is to use $(s,r)$ as the conditioning context to maximize the logit (equivalently, the likelihood) of the target object $o$. Prior studies have shown that FFNs are closely related to knowledge storage in Transformer models~\citep{geva2021transformer,dai2022knowledge}; accordingly, many studies achieve this goal by editing the FFN modules~\citep{meng2022locating,meng2022mass,fang2024alphaedit}. We first outline the model editing process and then formalize it in the context of GR.

\paragraph{Information Flow in FFN}
Following the notations in model editing for language models~\citep{fang2024alphaedit,meng2022mass}, we briefly summarize the setup as follows. In a transformer architecture, given a hidden state $h\in\mathbb{R}^{d_h}$ at the edited position, the FFN first computes an intermediate activation and then projects it back to the output space:
\begin{equation}
k=\sigma(W_{\mathrm{in}}h)\in\mathbb{R}^{d_0},
\qquad
z=W_{\mathrm{out}}k\in\mathbb{R}^{d_1},
\end{equation}
where $W_{\mathrm{in}}\in\mathbb{R}^{d_0\times d_h}$, $W_{\mathrm{out}}\in\mathbb{R}^{d_1\times d_0}$, and $\sigma(\cdot)$ is the element-wise activation. $k$ is referred to as the \emph{key} and $z$ as the \emph{value} in the key--value memory of FFNs.

\paragraph{Editing Targets}

Assume we have $m$ edit requests. For the $i$-th request, we obtain an activation (key) vector $k_i$ by running the model on the corresponding edit context and extracting the FFN intermediate activation at the subject position ($s_i$) from the selected layer. We denote this original FFN output by $z_i=W_{\mathrm{out}}k_i$. To define the desired edited output at this site, we solve for a minimal intervention $\delta_i$ on the FFN output that makes the overall model prefer the target object sequence associated with the edit request:
\begin{equation}
\delta_i=\argmin_{\delta}\ -\log p_{\theta}\!\left(o_i \,\middle|\, (s_i,r_i);\ z_i+\delta_i\right),
\label{eq:delta}
\end{equation}
where $(s_i,r_i)$ represents the
context of the edit request. $p_{\theta}(\cdot\mid \mathrm{context}_i;\ z_i+\delta_i)$ denotes the model’s output distribution when, at the chosen site, the FFN output is replaced by $z_i+\delta_i$ (all other computations remain unchanged). The goal of model editing is to encode such targeted changes in hidden states into the model parameters, i.e., to apply a structured update to $W_{\mathrm{out}}$. We then set the edited value vector as $z'_i \triangleq z_i+\delta_i$. Typically, to inject new knowledge into the model, we seek an  \emph{ideal update matrix} $\Delta W_{\mathrm{out}}^{*}$ such that, for any $i$-th request among the $m$ edit requests, the following constraint holds: 
\begin{equation}
\left(W_{\mathrm{out}} + \Delta W_{\mathrm{out}}^{*}\right)k_i = z'_i .
\end{equation}
For convenience, we refer to $\Delta W_{\mathrm{out}}$ as the \emph{estimated update matrix} that approximates $\Delta W^{*}_{\mathrm{out}}$, and set $\widehat{W}_{\mathrm{out}}\triangleq W_{\mathrm{out}}+\Delta W_{\mathrm{out}}$.

\paragraph{Preliminary in GR}
In the natural language setting, the goal is to edit the value $z$ at the subject position by solving for an intervention $\delta$ that maximizes the probability of the target output (i.e., the object), thereby yielding a corresponding parameter update $\Delta W_{\mathrm{out}}$. However, for sequential data in generative recommendation (GR), two challenges arise: (1) the lack of explicit sentence structure makes it difficult to construct an effective edit request $\langle s,r,o\rangle$; and (2) the GR target (e.g., cold-start item patterns) does not exhibit stable token co-occurrence, making a single edit that simultaneously improves all tokens in the pattern unreliable.
To address these issues, we treat a \emph{single token} as the object $o$ to circumvent the challenge posed by the lack of stable token co-occurrence; we define the \emph{entire item history together with the prefix of the cold-start item} as the subject $s$ and drop the relation $r$ to overcome the challenge of lack explicit subject. Formally, assume that each item is represented by a four-digit semantic ID (SID), with position $p\in\{0,1,2,3\}$. We denote the object token at position $p$ as $o_p$. Accordingly, each edit request of GR can be written in the following form: $\langle s_p, o_p \rangle$. These small modifications have far-reaching implications, which we detail in the following sections.

\section{\ourname: The Proposed Approach} 
This section introduces our proposed framework, \ourname, which adapts model editing to generative recommendation. As illustrated in Figure~\ref{fig:method}, \ourname consists of three components: \emph{Position-Wise Knowledge Preparation}, \emph{Locate-Then-Edit} framework, and \emph{One-One Triggering Policy}. We briefly summarize their roles as follows.
(1)~\textbf{Position-Wise Knowledge Preparation.} We construct pseudo interaction data for cold-start items and use it to form edit requests.
(2)~\textbf{Locate-Then-Edit Framework.} For each position, we first use a classifier to localize the key layer that is most strongly associated with that position, and then perform targeted edits on the identified layer.
(3)~\textbf{One-One Triggering Policy.} To prevent interference among edits for different positions, we adopt an adaptive triggering strategy at inference time, where the corresponding edit layer is triggered according to the current decoding position.



\subsection{\mbox{Position-Wise Knowledge Preparation}}
\label{sec:knowledge_preparation}

To address this challenge, we construct pseudo histories for cold-start items via similarity-based imputation, as illustrated on the left side of Figure~\ref{fig:method}. Concretely, for each cold-start item $c$, we first encode its available semantic information (often the only information accessible for cold-start items) into an embedding:
\begin{equation}
\bm{e}_c = f_{\mathrm{enc}}(\bm{m}_c),
\end{equation}
where $\bm{m}_c$ denotes the meta information of item $c$ and $f_{\mathrm{enc}}(\cdot)$ is an encoder\footnote{we use Sentence T5 as the encoder.}.

Next, we retrieve the top-$k$ most similar warm items based on embedding similarity. For each warm item $j \in \mathcal{I}_{\mathrm{warm}}$ with embedding $\bm{e}_j$, we compute
\begin{equation}
\mathrm{sim}(c,j) = \cos(\bm{e}_c, \bm{e}_j)
= \frac{\bm{e}_c^\top \bm{e}_j}{\lVert \bm{e}_c\rVert_2 \,\lVert \bm{e}_j\rVert_2},
\end{equation}
and select the top-$k$ warm items that are most similar to each cold-start item:
\begin{equation}
\mathcal{N}_k(c) = \mathrm{TopK}_{j \in \mathcal{I}_{\mathrm{warm}}}\ \mathrm{sim}(c,j).
\end{equation}
We use $\mathcal{N}_k(c)$ as a set of reference items to approximate how the cold-start item may be interacted with in real-world scenarios.

Finally, for each retrieved warm item $j \in \mathcal{N}_k(c)$, we extract the user interactions that precede $j$ in the user’s real interaction sequence and use these preceding interactions as a surrogate interaction history for $c$. By doing so, we obtain a set of synthesized, high-quality pseudo interaction sequences for cold-start items. We then split each target item by position to further construct position-wise edit requests, i.e., pseudo knowledge pairs $\langle s_p, o_p\rangle$, where $s_p$ consists of the interaction history together with the prefix of the target item, and $o_p$ is the target token at the corresponding position. These pairs can be directly used for model editing in GR.

\subsection{\mbox{Locate-Then-Edit Framework}}
\label{sec:locatethenedit}


As illustrated in the middle stage of Figure~\ref{fig:method}, we adopt a \emph{locate-then-edit} framework, which first identifies the most critical layers and then performs targeted updates. In this section, we describe the procedure from three aspects: (1) \textbf{Layer location}. \ourname injects the knowledge of cold-start items by updating a small subset of relevant layers. Hence, we aim to identify the layers that are most responsible for representing the target knowledge, so that we can introduce cold-start items while minimizing disruption to the model’s original recommendation capability. (2) \textbf{Memory construction}. We compute the key ($k$) and value ($z'$) of the cold-start items that the selected layers should store, i.e., the memory vectors that encode the desired behavioral changes. (3) \textbf{Parameter updating}. We then update parameters to store a portion of the constructed memories in each selected layer to achieve effective knowledge insertion with fewer side effects.

\subsubsection{Layer Location}
\label{sec:layer}

Recent studies have shown that editing FFN modules at different layers can affect model capabilities to varying degrees, making layer selection a critical design choice~\citep{geva2021transformer,dai2022knowledge}. Some prior work identifies a small set of ``most sensitive'' layers by measuring how perturbations to layer-wise hidden states influence the final outputs; however, such sensitivity-based selection can compromise the model’s original capabilities when edits are applied~\citep{li2023inference,fang2024alphaedit}. Our goal is to inject cold-start item knowledge while minimizing degradation of the model’s original recommendation performance. Notably, many model-editing methods can be viewed as learning a parameter update $\Delta W$ that linearly maps the activation (key) $k$ associated with both new and existing knowledge to their desired FFN outputs $z$. The success of such an update depends crucially on whether the activations (keys) corresponding to new knowledge are distinguishable from those of preserved knowledge in the target layer’s representation space. Therefore, our key idea is to train a linear probing classifier on the key to discriminate between positive and negative samples, following established probing frameworks~\citep{belinkov2022probing}.  

Because our synthesized data are position-wise, for each edit request at position $p$ we extract the activation (key) at the last token of the subject $s_p$ from each layer $l$, denoted as $k_{p}^{l}$, and assign it the label $1$. In addition, we apply the same procedure to samples from the original training set, collecting their corresponding key activations and assigning them the label $0$. This yields a probing dataset: $ \mathcal{D}^{(p,l)}=\{(k^{l}_{p,i}, y_i)\}_{i=1}^{n+m}$  where  $y_i=0$ if it comes from the original dataset (i.e., $i\in\{1,\ldots,N\}$), and $y_i=1$ if the sample comes from the $m$ edit requests (i.e., $i\in\{n+1,\ldots,n+m\}$).  To identify the most distinguishable layer for each position $p$, we define a linear classifier $Cls(k^{l}_{p,i}) = \mathrm{Sigmoid}(\langle \theta, k^{l}_{p,i}\rangle)$ to detect whether the activation (key) originates from cold-start items (edited knowledge) or from the original knowledge distribution. We split the dataset 8:2 into training and validation and train a binary linear classifier $Cls(\cdot)$ on the training split. For each position, we take the top-$1$ layer by accuracy as the edit layer denoted as $l_{p}$.

\subsubsection{Memory Construction}
\label{sec:memoryconstruction}

For each position in the edit requests, we determine an edit layer $l_p$ according to Section~\ref{sec:layer}, at which the new knowledge should be fully injected. Specifically, for the $i$-th edit request $(s_{p,i}, o_{p,i})$ of total $m$ edit requests, we compute a hidden vector $z'_i$ to replace the original output $z_i$ by the FFN of the layer $l_p$ such that adding $\delta_i = z'_i - z_i$ to the hidden state at layer $l_p$ and the generated token $T$ (i.e., the SID token of the cold-start item at position $p$) fully conveys the desired knowledge. 
For each edit request $\langle s_{p}, o_{p} \rangle_i$, to obtain the corresponding $\delta_i$ at the edit layer $l_p$, we optimize $\delta_i$ according to Eq.~\eqref{eq:delta} with the following loss:
\begin{equation}
\mathcal{L}_{\mathrm{edit}}(\delta_i)
=
\mathcal{L}_{\mathrm{CE}}\left(
\mathrm{onehot}(o_{p,i}),\ 
p_{\theta}\left(\cdot \mid s_{p,i};\ z_i+\delta_i\right)
\right),
\end{equation}
where $\mathrm{onehot}(\cdot)$ denotes the one-hot encoding of the token $o_{p,i}$, i.e., $\mathrm{onehot}(o_{p,i}) \in \mathbb{R}^{|\mathcal{V}|}$, where $|\mathcal{V}|$ is the vocabulary size. $\mathcal{L}_{\mathrm{CE}}$ denotes the cross-entropy loss. 




\subsubsection{Parameter Updating}
\label{sec:parameterupdate}
In generative recommendation, we treat \emph{cold-start items} as new knowledge to be injected into a pretrained model while preserving its performance on the warm items.

For the $m$ edit requests at position $p$, following Section~\ref{sec:memoryconstruction}, we obtain $\{k_i\}_{i=0}^{m-1}$, $\{z_i\}_{i=0}^{m-1}$, and $\{\delta_i\}_{i=0}^{m-1}$, where $z'_i = z_i + \delta_i$. If we stack keys and memories as matrices $K_1 = [k_1 \mid k_2 \mid \cdots \mid k_m]$ and
$Z'_1 = [z'_1 \mid z'_2 \mid \cdots \mid z'_n]$,
then the parameter update process could be optimizing by the following expanded least-squares objective:
\begin{equation}
W_1 \triangleq \argmin_{\widehat{W}}
\left(
\left\lVert \widehat{W}K_0 - Z_0\right\rVert_F^{2}
+
\left\lVert \widehat{W}K_1 - Z'_1\right\rVert_F^{2}
\right),
\end{equation}
where $(K_0,Z_0)$ are key--value  pairs extracted from the original training data, and $(K_1,Z'_1)$ are key--value pairs constructed for the edit requests. The first term enforces that the updated weights maintain the model's original mapping on $K_0$, thereby preserving the original recommendation capability on warm items, while the second term encourages the updated weights to realize the desired associations for cold-start items, thereby injecting new knowledge.

Applying the normal equation in block form yields
\begin{equation}
W_1\,[K_0\ \ K_1]\,[K_0\ \ K_1]^{\top}
=
[Z_0\ \ Z'_1]\,[K_0\ \ K_1]^{\top},
\end{equation}
which expands to
\begin{equation}
(W_0+\Delta W)\,(K_0K_0^{\top}+K_1K_1^{\top})
=
Z_0K_0^{\top}+Z'_1K_1^{\top}.
\end{equation}
Subtracting the pretraining normal equation $W_0K_0K_0^{\top}=Z_0K_0^{\top}$ gives the key trade-off relation:
\begin{equation}
\Delta W\,(K_0K_0^{\top}+K_1K_1^{\top})
=
Z'_1K_1^{\top}-W_0K_1K_1^{\top},
\end{equation}
showing that $\Delta W$ must correct the residual of the cold-start associations under the old weights, while being mediated by both the covariance of keys from the original training data ($K_0K_0^{\top}$, reflecting preservation pressure) and the covariance of keys from cold-start items ($K_1K_1^{\top}$, reflecting injection pressure). Defining $C_0\triangleq K_0K_0^{\top}$ and $R\triangleq Z'_1-W_0K_1$, we obtain the closed-form update:
\begin{equation}
\Delta W = R K_1^{\top}\left(\lambda C_0 + K_1K_1^{\top}\right)^{-1},
\label{eq:updateW}
\end{equation}
where $\lambda$ controls the preservation--injection trade-off.

\subsection{\mbox{One-One Triggering Policy in Inference }}
\label{sec:one_one}

To accommodate the unique data characteristics in GR (i.e., the absence of stable token bundles), we adopt a position-wise knowledge preparation and editing scheme as illustrated above. Under this setting, a new challenge emerges at inference time: generating a single item typically requires producing multiple SID tokens (four digits in \ourname). Since we perform edits for each position $p$ on a specific edit layer $l_p$, our editing procedure is optimized to affect only the token at position $p$ (i.e., $o_p$), without explicitly accounting for its potential influence on tokens at other positions. However, following prior practice~\citep{meng2022mass}, the edits on all designated layers (i.e., $\{l_p\}_{p\in[0,3]}$) would otherwise be simultaneously triggered during decoding. Such concurrent triggering can introduce uncontrolled cross-position interactions, leading to unpredictable token outputs and undermining the reliability of position-wise editing.

To address this issue, we propose a \emph{One-One Triggering Policy}. As illustrated in Figure~\ref{fig:method} (Right), we introduce a gating mechanism that enforces edit-inference alignment: when generating the SID token at position $p$, the forward pass triggers only the edit associated with layer $l_p$, while the edits for all other positions remain inactive. This One-One activation prevents inter-position coupling effects, ensuring that each position is influenced exclusively by its intended edit and thereby stabilizing multi-token SID generation in GR.

To provide robustness analyses for the One-One Triggering policy, we demonstrate that it enables low-interference injection of cold-start SIDs by bounding cross-layer interference.

\begin{assumption}[Bounded Cross-Layer Interference]
\label{assump:1}
For edit layers $l_p$ corresponding to different SID positions $p \in \{0,1,2,3\}$, the cross-layer contributions from non-target updates are bounded, i.e., for any $q \neq p$, there exists a small positive interference threshold $\varepsilon > 0$ such that $\|\Delta W_q \cdot k_p\|_2 \leq \varepsilon$, where $k_p = \sigma(W_{\mathrm{in}}^{(l_p)} h_S)$ is the FFN input activation of layer $l_p$ when generating the token $o_p$.
\end{assumption}

\begin{theorem}
\label{thm:core}
Under Assumption~\ref{assump:1}, let $z_p^* \triangleq (W_{\mathrm{out}}^{(l_p)} + \Delta W_p) \cdot k_p$ be the ideal FFN output of layer $l_p$ targeting token $o_p$, $p_{\mathrm{ideal}} \triangleq \mathrm{softmax}(z_p^*)_{o_p}$ be the ideal probability where only $\Delta W_p$ is applied using one-one triggering policy in GenRecEdit  (i.e., $\Delta W_q$ for $q \neq p$ are inactive at layer $l_p$), and the interference threshold defined in Assumption~\ref{assump:1} satisfies $\varepsilon < \frac{1}{6} ( \langle z_p^*, e_{o_p} \rangle - \max_{o \neq o_p} \langle z_p^*, e_o \rangle )$, then the cross-token interference (i.e., the cumulative impact of $\Delta W_q$ for $q \neq p$ on the generation of $o_p$) is negligible. Specifically, the actual generation probability $p_{\mathrm{actual}}$ satisfies:
$$
\left| p_{\mathrm{actual}}(o_p) - p_{\mathrm{ideal}}(o_p) \right| \leq \delta(\varepsilon),
$$
where $\delta(\varepsilon) \triangleq \frac{1}{4}(e^{6\varepsilon}-1) \approx 1.5\varepsilon$, ensuring the robustness of one-one triggering policy in GenRecEdit.
\end{theorem}
See Appendix~A for proof details.

\section{Experiments}

We conducted comprehensive experiments on three Amazon datasets. The results and analyses demonstrate the effectiveness of \ourname in injecting cold-start item patterns. 
The code is available\footnote{https://github.com/Starrylay/GenRecEdit.git}.


\subsubsection{Dataset}
We evaluate our method on three categories from the widely used Amazon 2023 Review~\footnote{\url{https://jmcauley.ucsd.edu/data/amazon/}} dataset~\cite{hou2024bridging}: \emph{Video~Games}, \emph{Software}, and \emph{Cell Phones and Accessories}. Following~\cite{TIGER,VQRec}, we treat each user’s historical reviews as interaction records and sort them chronologically, with the earliest review placed first. For data processing, we adopt the standard timestamp-based protocol~\cite{SASREC,TIGER,ding2026inductive}: we use the official dataset splits that are strictly partitioned into training, validation, and test sets by time, with no temporal overlap. On top of this protocol, we remove cold-start users from the training data to avoid confounding effects when analyzing cold-start items. Specifically, we collect the set of users appearing in the test split and filter the training and validation splits by retaining only interactions from these users. For evaluation, to clearly quantify performance on cold-start items, we further partition the test set based on whether the target item appears in the training split, resulting in a cold subset, a warm-item test set, and an overall test set that contains both. Across the three datasets, the cold subset accounts for $65.6\%$, $24.3\%$, and $75.5\%$ of the overall test set, respectively.



\begin{table*}[t!]
\small
\centering
\caption{
The overall recommendation performance of various methods across the three datasets (Overall) and the cold subsetresults (Cold). 
The best-performing and second-best methods in Overall are denoted with boldface and underlining, respectively. The ``-'' indicates that the item ID-based method cannot handle cold-start items, and we could treat its value as $0$. The improvements over the second-best methods are statistically significant (paired $t$-test, $p$-value$<0.05$).
}
\label{tab:main_result_overall_cold}
\setlength{\tabcolsep}{2.8pt}
\renewcommand{\arraystretch}{1.12}
\resizebox{1.0\linewidth}{!}{
\begin{tabular}{l *{18}{c}}
\toprule
\multirow{3}{*}{Metric}
& \multicolumn{4}{c}{Item ID-based}
& \multicolumn{6}{c}{Semantic ID-based}
& \multicolumn{8}{c}{Cold-start-based} \\
\cmidrule(lr){2-5}\cmidrule(lr){6-11}\cmidrule(lr){12-19}
& \multicolumn{2}{c}{SASRec} & \multicolumn{2}{c}{BERT4Rec}
& \multicolumn{2}{c}{VQ-Rec} & \multicolumn{2}{c}{TIGER} & \multicolumn{2}{c}{LC-Rec}
& \multicolumn{2}{c}{Retrain} & \multicolumn{2}{c}{Finetune} & \multicolumn{2}{c}{SpecGR} & \multicolumn{2}{c}{\textbf{Ours}} \\
\cmidrule(lr){2-3}\cmidrule(lr){4-5}
\cmidrule(lr){6-7}\cmidrule(lr){8-9}\cmidrule(lr){10-11}
\cmidrule(lr){12-13}\cmidrule(lr){14-15}\cmidrule(lr){16-17}\cmidrule(lr){18-19}
& Overall & \textcolor{coldtxt}{Cold} & Overall & \textcolor{coldtxt}{Cold}
& Overall & \textcolor{coldtxt}{Cold} & Overall & \textcolor{coldtxt}{Cold} & Overall & \textcolor{coldtxt}{Cold}
& Overall & \textcolor{coldtxt}{Cold} & Overall & \textcolor{coldtxt}{Cold} & Overall & \textcolor{coldtxt}{Cold} & Overall & \textcolor{coldtxt}{Cold} \\
\midrule

\rowcolor{gray!10}
\multicolumn{19}{l}{\textit{Video}} \\
NDCG@10
& 0.0014 & \coldcell{--} & 0.0019 & \coldcell{--}
& 0.0027 & \coldcell{0.0000} & 0.0070 & \coldcell{0.0020} & 0.0055 & \coldcell{0.0012}
& 0.0083 & \coldcell{0.0044} & 0.0076 & \coldcell{\underline{0.0103}} & \underline{0.0114} & \coldcell{0.0051}
& \cellcolor{gray!15}\textbf{0.0118} & \cellcolor{gray!15}\coldcell{\textbf{0.0123}} \\
NDCG@20
& 0.0019 & \coldcell{--} & 0.0028 & \coldcell{--}
& 0.0035 & \coldcell{0.0000} & 0.0096 & \coldcell{0.0030} & 0.0072 & \coldcell{0.0017}
& 0.0109 & \coldcell{0.0058} & 0.0093 & \coldcell{\underline{0.0125}} & \underline{0.0136} & \coldcell{0.0062}
& \cellcolor{gray!15}\textbf{0.0140} & \cellcolor{gray!15}\coldcell{\textbf{0.0141}} \\
NDCG@50
& 0.0027 & \coldcell{--} & 0.0044 & \coldcell{--}
& 0.0046 & \coldcell{0.0001} & 0.0137 & \coldcell{0.0052} & 0.0099 & \coldcell{0.0028}
& \underline{0.0158} & \coldcell{0.0096} & 0.0124 & \coldcell{\underline{0.0164}} & \underline{0.0158} & \coldcell{0.0075}
& \cellcolor{gray!15}\textbf{0.0182} & \cellcolor{gray!15}\coldcell{\textbf{0.0176}} \\
RECALL@10
& 0.0028 & \coldcell{--} & 0.0046 & \coldcell{--}
& 0.0053 & \coldcell{0.0000} & 0.0141 & \coldcell{0.0043} & 0.0109 & \coldcell{0.0026}
& 0.0167 & \coldcell{0.0094} & 0.0141 & \coldcell{\underline{0.0191}} & \textbf{0.0237} & \coldcell{0.0106}
& \cellcolor{gray!15}\underline{0.0210} & \cellcolor{gray!15}\coldcell{\textbf{0.0209}} \\
RECALL@20
& 0.0051 & \coldcell{--} & 0.0081 & \coldcell{--}
& 0.0085 & \coldcell{0.0001} & 0.0243 & \coldcell{0.0083} & 0.0177 & \coldcell{0.0045}
& 0.0270 & \coldcell{0.0148} & 0.0209 & \coldcell{\underline{0.0281}} & \textbf{0.0409} & \coldcell{0.0195}
& \cellcolor{gray!15}\underline{0.0299} & \cellcolor{gray!15}\coldcell{\textbf{0.0283}} \\
RECALL@50
& 0.0089 & \coldcell{--} & 0.0163 & \coldcell{--}
& 0.0141 & \coldcell{0.0006} & 0.0455 & \coldcell{0.0197} & 0.0318 & \coldcell{0.0102}
& \underline{0.0521} & \coldcell{0.0344} & 0.0364 & \coldcell{\underline{0.0475}} & \textbf{0.0953} & \coldcell{\textbf{0.0504}}
& \cellcolor{gray!15}0.0510 & \cellcolor{gray!15}\coldcell{0.0457} \\
\addlinespace[2pt]
\midrule
\rowcolor{gray!10}
\multicolumn{19}{l}{\textit{Software}} \\
NDCG@10
& 0.0321 & \coldcell{--} & 0.0308 & \coldcell{--}
& 0.0263 & \coldcell{0.0002} & \underline{0.0353} & \coldcell{0.0005} & 0.0351 & \coldcell{0.0016}
& 0.0350 & \coldcell{0.0008} & 0.0347 & \coldcell{\underline{0.0303}} & 0.0328 & \coldcell{0.0181}
& \cellcolor{gray!15}\textbf{0.0370} & \cellcolor{gray!15}\coldcell{\textbf{0.0228}} \\
NDCG@20
& 0.0412 & \coldcell{--} & 0.0412 & \coldcell{--}
& 0.0333 & \coldcell{0.0006} & 0.0451 & \coldcell{0.0005} & \underline{0.0456} & \coldcell{0.0017}
& 0.0455 & \coldcell{0.0009} & 0.0437 & \coldcell{\underline{0.0318}} & 0.0381 & \coldcell{0.0222}
& \cellcolor{gray!15}\textbf{0.0475} & \cellcolor{gray!15}\coldcell{\textbf{0.0249}} \\
NDCG@50
& 0.0522 & \coldcell{--} & 0.0580 & \coldcell{--}
& 0.0442 & \coldcell{0.0017} & \underline{0.0604} & \coldcell{0.0013} & 0.0603 & \coldcell{0.0024}
& \textbf{0.0606} & \coldcell{\textbf{0.0552}} & 0.0552 & \coldcell{\underline{0.0357}} & 0.0447 & \coldcell{0.0285}
& \cellcolor{gray!15}\underline{0.0601} & \cellcolor{gray!15}\coldcell{\underline{0.0286}} \\
RECALL@10
& 0.0673 & \coldcell{--} & 0.0622 & \coldcell{--}
& 0.0493 & \coldcell{0.0004} & \underline{0.0719} & \coldcell{0.0013} & \textbf{0.0720} & \coldcell{0.0031}
& 0.0678 & \coldcell{0.0013} & 0.0641 & \coldcell{\textbf{0.0418}} & 0.0549 & \coldcell{0.0311}
& \cellcolor{gray!15}0.0706 & \cellcolor{gray!15}\coldcell{\underline{0.0359}} \\
RECALL@20
& 0.1039 & \coldcell{--} & 0.1038 & \coldcell{--}
& 0.0770 & \coldcell{0.0018} & \underline{0.1111} & \coldcell{0.0013} & 0.1100 & \coldcell{0.0036}
& 0.1094 & \coldcell{0.0018} & 0.1003 & \coldcell{\underline{0.0476}} & 0.0770 & \coldcell{\textbf{0.0480}}
& \cellcolor{gray!15}\textbf{0.1122} & \cellcolor{gray!15}\coldcell{0.0445} \\
RECALL@50
& 0.1587 & \coldcell{--} & 0.1789 & \coldcell{--}
& 0.1326 & \coldcell{0.0081} & \textbf{0.1882} & \coldcell{0.0054} & \textbf{0.1882} & \coldcell{0.0067}
& \underline{0.1860} & \coldcell{0.0013} & 0.1575 & \coldcell{\underline{0.0678}} & 0.1107 & \coldcell{\textbf{0.0792}}
& \cellcolor{gray!15}0.1754 & \cellcolor{gray!15}\coldcell{0.0629} \\
\addlinespace[2pt]
\midrule

\rowcolor{gray!10}
\multicolumn{19}{l}{\textit{Phone}} \\
NDCG@10
& 0.0006 & \coldcell{--} & 0.0010 & \coldcell{--}
& 0.0005 & \coldcell{0.0000} & 0.0037 & \coldcell{0.0014} & 0.0028 & \coldcell{0.0004}
& \underline{0.0045} & \coldcell{0.0033} & 0.0035 & \coldcell{\underline{0.0041}} & 0.0030 & \coldcell{0.0018}
& \cellcolor{gray!15}\textbf{0.0064} & \cellcolor{gray!15}\coldcell{\textbf{0.0052}} \\
NDCG@20
& 0.0010 & \coldcell{--} & 0.0013 & \coldcell{--}
& 0.0006 & \coldcell{0.0000} & 0.0050 & \coldcell{0.0020} & 0.0036 & \coldcell{0.0006}
& \underline{0.0057} & \coldcell{0.0042} & 0.0044 & \coldcell{\underline{0.0052}} & 0.0044 & \coldcell{0.0032}
& \cellcolor{gray!15}\textbf{0.0078} & \cellcolor{gray!15}\coldcell{\textbf{0.0063}} \\
NDCG@50
& 0.0016 & \coldcell{--} & 0.0019 & \coldcell{--}
& 0.0009 & \coldcell{0.0001} & 0.0070 & \coldcell{0.0029} & 0.0049 & \coldcell{0.0009}
& \underline{0.0078} & \coldcell{0.0058} & 0.0055 & \coldcell{\underline{0.0066}} & 0.0057 & \coldcell{0.0042}
& \cellcolor{gray!15}\textbf{0.0098} & \cellcolor{gray!15}\coldcell{\textbf{0.0079}} \\
RECALL@10
& 0.0012 & \coldcell{--} & 0.0020 & \coldcell{--}
& 0.0012 & \coldcell{0.0001} & 0.0072 & \coldcell{0.0026} & 0.0057 & \coldcell{0.0009}
& 0.0081 & \coldcell{0.0055} & 0.0062 & \coldcell{\underline{0.0075}} & \underline{0.0102} & \coldcell{0.0068}
& \cellcolor{gray!15}\textbf{0.0108} & \cellcolor{gray!15}\coldcell{\textbf{0.0083}} \\
RECALL@20
& 0.0028 & \coldcell{--} & 0.0032 & \coldcell{--}
& 0.0018 & \coldcell{0.0002} & 0.0124 & \coldcell{0.0048} & 0.0088 & \coldcell{0.0015}
& 0.0129 & \coldcell{0.0088} & 0.0099 & \coldcell{0.0118} & \textbf{0.0204} & \coldcell{\textbf{0.0144}}
& \cellcolor{gray!15}\underline{0.0165} & \cellcolor{gray!15}\coldcell{\underline{0.0126}} \\
RECALL@50
& 0.0059 & \coldcell{--} & 0.0062 & \coldcell{--}
& 0.0031 & \coldcell{0.0003} & 0.0228 & \coldcell{0.0099} & 0.0156 & \coldcell{0.0032}
& 0.0234 & \coldcell{0.0171} & 0.0157 & \coldcell{0.0188} & \textbf{0.0369} & \coldcell{\textbf{0.0245}}
& \cellcolor{gray!15}\underline{0.0265} & \cellcolor{gray!15}\coldcell{\underline{0.0207}} \\
\bottomrule
\end{tabular}
}
\end{table*}

\subsubsection{Baselines}
We first compared our approach with traditional item ID-based methods.
\textbf{Item ID-based}:
(1)~\textbf{SASRec}~\cite{SASREC} employs a unidirectional Transformer to capture sequential dependencies;  
(2)~\textbf{BERT4Rec}~\cite{BERT4REC} utilizes a bidirectional Transformer trained with a cloze-style objective;   
We also compared our approach with generative recommendation methods based on semantic IDs.
\textbf{Semantic ID-based}:
(3)~\textbf{VQ-Rec}~\cite{VQRec} applies product quantization to tokenize items into semantic IDs, which are then pooled to obtain item representations;
(4)~\textbf{TIGER}~\cite{TIGER} utilizes RQ‑VAE to generate codebook identifiers, embedding semantic information into discrete code sequences;
(5)~\textbf{LC-Rec}~\cite{LC_Rec} exploits identifiers with auxiliary alignment tasks to associate the generated codes with natural language;
Finally, we compare our approach against baselines designed to handle cold-start items:
(6)~\textbf{Retraining} retrains the model from scratch using the full training data augmented with the synthesized cold-start interactions.
(7)~\textbf{Finetuning} continues training a well-trained model using only the synthesized cold-start interactions.
(8)~\textbf{SpecGR}~\cite{ding2026inductive} is a plug-and-play framework for inductive generative recommendation that drafts candidate items and uses the GR model to verify them.

\subsubsection{Evaluation Metrics}
Following prior studies~\cite{TIGER,LC_Rec}, we evaluate performance using two commonly adopted ranking metrics: top-$k$ \textit{Recall} and top-$k$ \textit{Normalized Discounted Cumulative Gain} (NDCG).  We report results for $k \in \{10, 20, 50\}$. In addition, as illustrated in Section~\ref{sec:analysis}, we include \textit{IID Ratio} as an evaluation metric to provide an intuitive measure of how well the model captures cold-start item patterns. For example, on the cold subset, a higher \textit{IID Ratio} indicates that a larger fraction of the recommended items are cold-start items, and vice versa. Similarly, on the warm-start test set, a higher \textit{IID Ratio} indicates that a larger fraction of the recommended items are warm-start items, and vice versa.

\subsubsection{Implementation Details}

For the tokenization module, we adopt Sentence-T5~\cite{ni2021sentence} to encode each item’s title and other textual metadata into embeddings. Following Tiger~\citep{TIGER}, we use M=3 codebooks, each containing K=256 code vectors of dimensionality d=32. However, we do not append an additional token to the end of the ordered semantic codes, as doing so may disrupt the semantic coherence among the SIDs within an item and thereby increase the difficulty of editing. The weighting coefficient $\lambda$ in Eq.~\eqref{eq:updateW} is analyzed in our analysis experiments, and we set $\lambda \in \{3{,}000,\, 3{,}000,\, 1{,}000\}$ for the three datasets, respectively, which yields the best performance. We remove users from the training set who do not appear in the test set, to eliminate the confounding effects of cold-start users in evaluation. \ourname is built upon T5~\footnote{\url{https://huggingface.co/docs/transformers/en/model_doc/t5}}. We set the number of decoder layers to 6 and use \texttt{gated-silu} as the FFN activation, which is crucial for successful editing because the default ReLU activation can induce strong truncation, leading to low-rank key representations $k$ and, consequently, making the matrix inversion in Eq.~\eqref{eq:updateW} ill-conditioned. The multi-head RQ-VAE is trained for 10{,}000 epochs using the Adam optimizer~\cite{loshchilov2017decoupled} with a learning rate of $1e^{-3}$ and a batch size of 2{,}048.

\subsection{Overall Performances}

Table~\ref{tab:main_result_overall_cold} reports the results on both the overall test set and the cold subset across the three datasets. We draw the following three key conclusions regarding performance on the overall test set. 
\textbf{(1)} \ourname generally achieves the best overall performance across multiple metrics (including NDCG and Recall) compared with both conventional item ID-based methods and generative semantic ID-based methods. In addition, \ourname remains competitive with representative cold-start baselines, supporting the feasibility of applying model editing to generative recommendation. 
\textbf{(2)} Semantic ID-based generative recommenders consistently outperform traditional item ID-based methods. A key reason is that semantic IDs provide a content-informed discrete representation that can be assigned to previously unseen items (cold-start items), enabling the model to reason about cold-start items through their semantic tokens. In contrast, ID-based recommenders rely on item-specific parameters (e.g., ID embeddings) learned from historical interactions; when an item is unseen, these parameters are missing or poorly estimated, which substantially weakens their ability to recommend cold-start items and degrades overall performance.
\textbf{(3)} \ourname, together with  \texttt{Retrain}, \texttt{Finetune}, and \texttt{SpecGR}, all those cold-start-specialized methods attains promising results. This suggests that the constructed interaction sequences targeting cold-start items provide effective supervisory signals, although different methods exploit them in different ways. Importantly, \ourname is substantially more efficient than these alternatives, which we analyze in detail in subsequent diagnostic experiments.

We further draw the following conclusions regarding performance on the cold subset. 
\textbf{(1)} Compared with both item ID-based and semantic ID-based models, \ourname delivers a clear improvement on cold-start items, indicating that our method successfully injects cold-start item knowledge and further demonstrating the viability of model editing for generative recommendation. 
\textbf{(2)} On the cold-start split of the \texttt{Software} dataset, our method still achieves comparable performance, whereas continued training (i.e., Finetune) tends to overfit to the newly injected data (high performance on the cold subset) and consequently induces catastrophic forgetting of prior knowledge (low performance on the warm subset and the overall test set). In fact, the trade-off between the injected data and the prior knowledge can be controlled by a hyperparameter of \ourname, which we discuss in the analysis experiments.

\begin{table}[!hb]
\small

\centering
\caption{Recommendation performance on the warm subset. Using the Phone dataset as an example. ``IID R.'' denotes IID Ratio; ``N.'' denotes NDCG; ``R.'' denotes RECALL; and ``Drop'' denotes the relative drop of N.@10 (\%).  }
\label{tab:phone_warm_small}

\renewcommand{\arraystretch}{1.12}
\setlength{\tabcolsep}{4.2pt}
\resizebox{1.0\columnwidth}{!}{
\begin{tabular}{lccccc c}
\toprule
\textbf{Method} & \textbf{IID R.@10} & \textbf{N.@10} & \textbf{N.@20} & \textbf{R.@10} & \textbf{R.@20} & \textbf{Drop} \\
\midrule


\rowcolor{gray!10}
\multicolumn{7}{l}{\textit{Semantic ID-based}} \\
TIGER  & 0.7020 & 0.0108 & 0.0144 & 0.0215 & 0.0362 & -- \\

\rowcolor{gray!10}
\multicolumn{7}{l}{\textit{Cold-start-based}} \\
Retraining & 0.6181 & 0.0080 & 0.0104 & 0.0165 & 0.0259 & -25.9\% \\
Finetuning & 0.1052 & 0.0014 & 0.0018 & 0.0022 & 0.0037 & -87.0\%\\
SpecGR     & 0.3329 & 0.0067 & 0.0081 & 0.0207 & 0.0389 &  -38.0\% \\
\rowcolor{gray!15}
\textbf{\ourname}
& \textbf{0.6366} & \textbf{0.0101} & \textbf{0.0127} & \textbf{0.0186} & \textbf{0.0288}
& \textbf{-6.5\%} \\
\bottomrule
\end{tabular}
}
\label{tab:warm}
\end{table}

Finally, we examine the performance of several cold-start-based methods on the warm subset. As shown in Table~\ref{tab:warm}, under the TIGER backbone, \ourname incurs the smallest loss on the warm subset in terms of NDCG@10, with only a $6.5\%$ drop.

\subsection{Ablation Study}
We conducted ablation studies to verify the effect of each module in \ourname. Specifically, we consider the following experimental settings: \textbf{(1) Knowledge preparation.} In Section~\ref{sec:knowledge_preparation}, we identify a key challenge of applying model editing to GR data: the lack of stable token bundles, which hinders controlling multi-token outputs with a single edit. We therefore propose a position-wise knowledge preparation scheme, and compare it with a conventional object-wise strategy, denoted as \texttt{w/o position-wise}. \textbf{(2) Layer Location.} In Section~\ref{sec:locatethenedit}, we locate critical layers via probing: for each position, we train a classifier at every layer that takes the key activation $k$ of the subject $s_p$ (i.e., the key at the last token of the interaction history and the prefix SIDs) and predicts whether $s_p$ belong to a cold-start item. We validate this classifier-based selection against two baselines that edit a \emph{random} layer or the layer with the \emph{lowest} classifier accuracy, denoted as \texttt{w/o classifier (random)} and \texttt{w/o classifier (worst)}. \textbf{(3) Inference.} As discussed in Section~\ref{sec:one_one}, we use the One-One Triggering Policy at inference to avoid interference among position-wise edits. We compare it with a conventional strategy~\citep{meng2022locating,meng2022mass} that keeps all edited layers continuously active so all the edits affect every decoded token, denoted as \texttt{w/o one-one triggering}.

As shown in Table~\ref{tab:ablation}, we draw two conclusions. (1) Each module and mechanism contributes positively to recommendation performance. (2) Among them, position-wise knowledge preparation and the One-One Triggering Policy are essential for performance, which aligns with the model editing challenges posed by GR data discussed in Section~\ref{sec:intro} and highlights the necessity and effectiveness of our~design.

\subsection{Experimental Analyses}

\begin{table}[t]
\small
\centering
\caption{
The results of the ablation study. Using the Phone dataset as an example. ``w/o'' indicates that the corresponding module is removed.
}
\vspace{-0.2cm}
\label{tab:ablation}
\renewcommand{\arraystretch}{1.15}
\setlength{\tabcolsep}{5pt}
\resizebox{1.0\columnwidth}{!}{
\begin{tabular}{lccccc}
\toprule
\textbf{Model} & \textbf{IID R.@10} & \textbf{N.@10} & \textbf{N.@20} & \textbf{R.@10} & \textbf{R.@20} \\
\midrule

\rowcolor{gray!15}
\textbf{\ourname} 
& \textbf{0.7359} & \textbf{0.0064} & \textbf{0.0078} & \textbf{0.0108} & \textbf{0.0165} \\
\midrule

\rowcolor{gray!10}
\multicolumn{6}{l}{\textit{Knowledge Preparation}} \\
w/o position-wise 
& 0.0220 & 0.0002 & 0.0002 & 0.0004 & 0.0005 \\
\midrule

\rowcolor{gray!10}
\multicolumn{6}{l}{\textit{Layer Location}} \\
w/o classifier (random) 
& 0.7339 & 0.0059 & 0.0072 & 0.0100 & 0.0150 \\
w/o classifier (worst)  
& 0.6865 & 0.0055 & 0.0066 & 0.0094 & 0.0137 \\
\midrule

\rowcolor{gray!10}
\multicolumn{6}{l}{\textit{Inference}} \\
w/o one-one triggering 
& 0.0030 & 0.0000 & 0.0001 & 0.0001 & 0.0002 \\
\bottomrule
\end{tabular}
}
\vspace{-0.4cm}
\end{table}

We further conduct experimental analyses to assess the effectiveness of \ourname. Unless otherwise specified, all experiments are conducted on the Video dataset.

\subsubsection{Sensitive Analysis of Knowledge Preparation}

New knowledge (comprising cold-start items and their synthesized interaction histories) is a central component of \ourname. Both the \emph{quality} and \emph{quantity} of this new knowledge are crucial for effective model editing. Accordingly, we conduct two experiments. 

\begin{figure}[th]
\centering
\includegraphics[width=1.0\columnwidth]{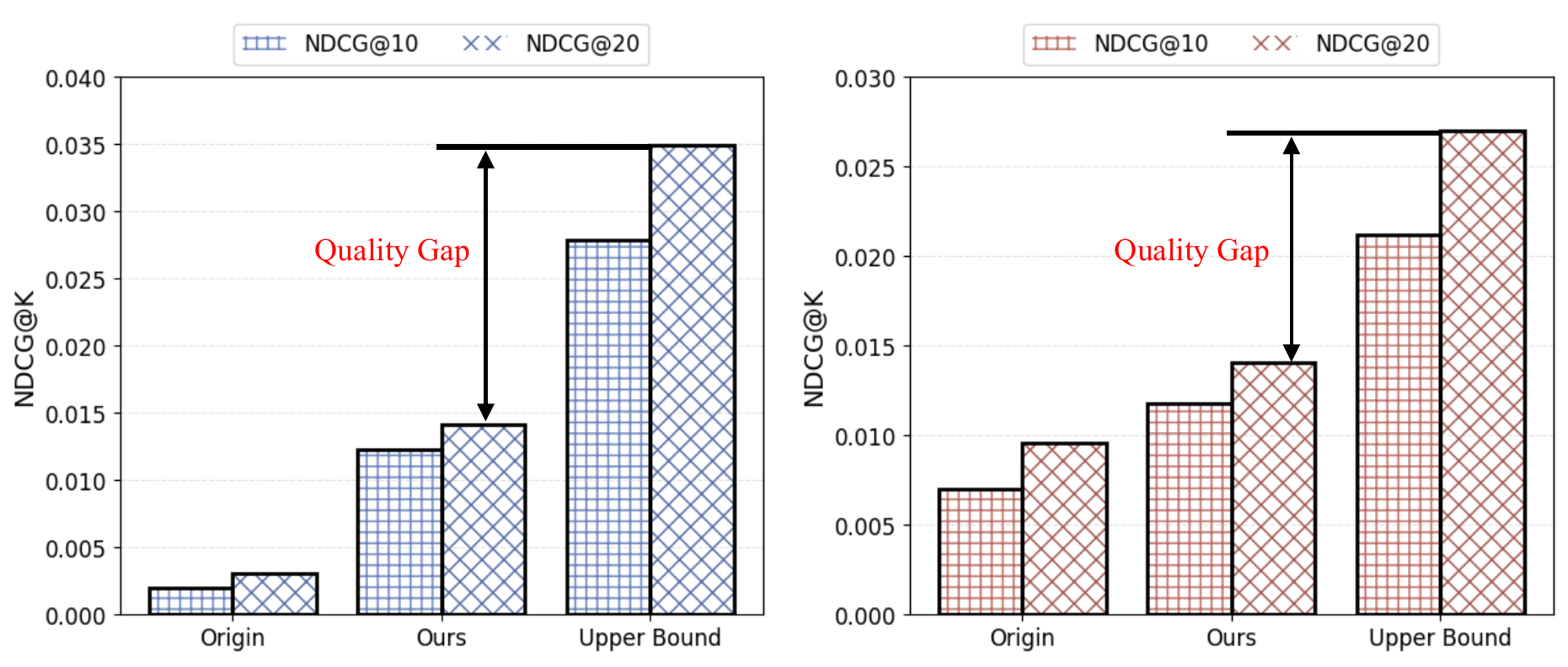}
\caption{
An analysis of the quality of constructed knowledge. The \textbf{left} panel shows performance under three settings on the cold subset, while the \textbf{right} panel reports the corresponding results on the overall test set. 
}

\label{fig:knowledgequality}
\vspace{-0.3cm}
\end{figure}

\textbf{(1) Quality of new knowledge.}
We construct two control settings: (a) no injection of new knowledge, denoted as \texttt{Origin}; and 
(b) injecting the ground-truth interaction histories of cold-start items as new knowledge during editing, denoted as \texttt{Upper Bound}. As shown in Figure~\ref{fig:knowledgequality}, we obtain the following observations. (a) On the cold subset, \ourname achieves a substantial improvement over \texttt{Origin}, demonstrating the effectiveness of injecting new knowledge. (b) Compared with \texttt{Upper Bound}, \ourname still exhibits a considerable performance margin, which we refer to as the Quality Gap.  This gap indicates significant headroom for better approximating the real interaction histories of cold-start items, leaving ample room for future research. We draw similar conclusions on the overall test set.

\begin{figure}[t]
\centering
\includegraphics[width=1.0\columnwidth]{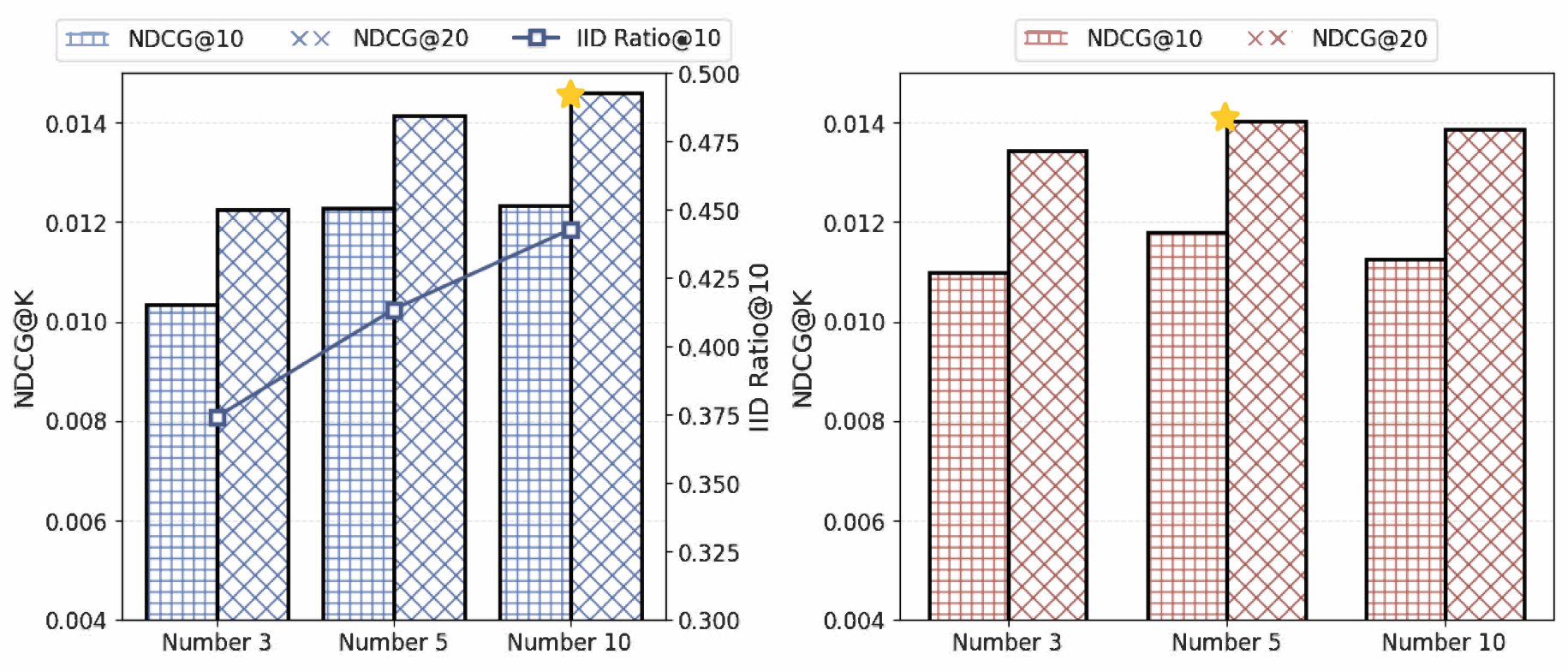}
\caption{
A sensitivity analysis on the number of constructed knowledge. The \textbf{left} figure shows changes of three metrics on the cold subset (NDCG@10, NDCG@20, and IID Ratio@10), while the \textbf{right} figure reports metric variations on the overall test set (NDCG@10 and NDCG@20).
}
\label{fig:knowledgenumber}
\end{figure}

\textbf{(2) Quantity of new knowledge.} We evaluate model performance under varying amounts of constructed knowledge injected per cold-start item. As shown in Figure~\ref{fig:knowledgenumber}, we draw three key findings. 
(a) As the injected knowledge per cold-start item increases, \ourname achieves a higher $IID~Ratio@10$ on the cold subset, indicating improved capture of the SID patterns of cold-start items.  (b) With more injected knowledge, recommendation accuracy for cold-start items (measured by NDCG) improves steadily. (c) On the full test set, overall recommendation accuracy first increases and then declines as the injected knowledge grows, suggesting an inherent trade-off between optimizing performance on warm items and on cold-start items.

\begin{figure}[t]
\centering
\includegraphics[width=1.0\columnwidth]{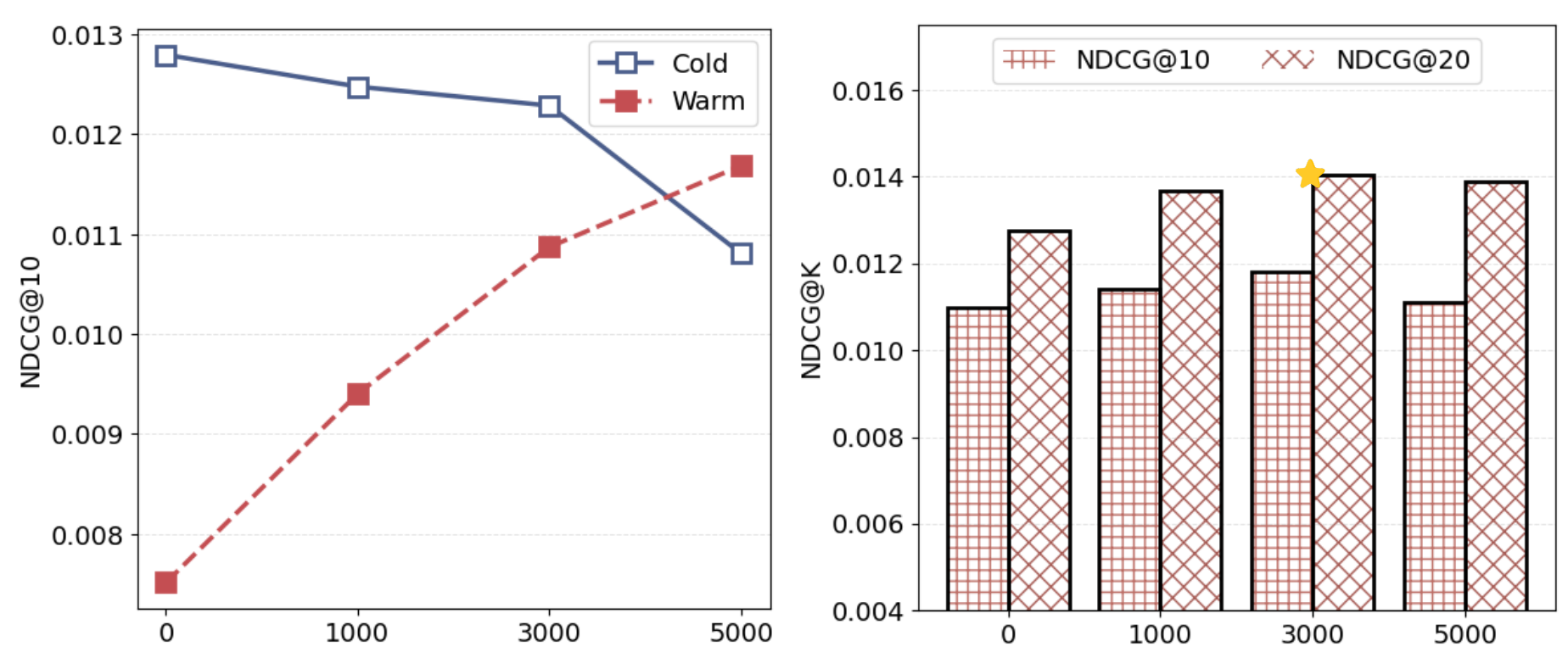}
\caption{
An analysis on the hyper parameter $\lambda$. The \textbf{left} figure shows the performance trends on cold-start and warm items (NDCG@10), while the \textbf{right} reports the performance on the overall test set (NDCG@10 and NDCG@20).
}
\label{fig:cov}
\end{figure}


\subsubsection{Sensitive Analysis of Hyper-parameters}

$\lambda$ in Eq.~\eqref{eq:updateW} is an important hyperparameter that governs the trade-off between preserving original knowledge (warm items) and injecting new knowledge (cold-start items). Therefore, an appropriate choice of $\lambda$ is essential for maximizing the utility of \ourname on the overall test set. In particular, a larger $\lambda$ enforces stronger preservation of the original knowledge, but correspondingly weakens the injection strength for cold-start items.  We sweep $\lambda$ over the range $[0, 5000]$ by evaluating four representative values. The results are reported in Figure~\ref{fig:cov}, from which we draw two key observations. \textbf{ (1)} As shown in the left panel, increasing $\lambda$ consistently improves performance on prior knowledge (warm items) while degrading performance on new knowledge (cold-start items), empirically validating the role of $\lambda$ in Eq.~\eqref{eq:updateW}. \textbf{(2)} The right part shows that a moderate increase in $\lambda$ can improve performance on the overall test set by better balancing the trade-off between warm and cold-start items.

\begin{table}[t]
\small
\centering
\caption{Comparison of update time across four algorithms.}
\vspace{-0.2cm}
\label{tab:update_time_compare}
\renewcommand{\arraystretch}{1.12}
\setlength{\tabcolsep}{6pt}
\resizebox{1.0\columnwidth}{!}{
\begin{tabular}{lcccc}
\toprule
\textbf{Model Update} & \textbf{Retraining} & \textbf{Fintuning} & \textbf{SpecGR} & \textbf{\ourname} \\
\midrule

\textbf{Type} & Train & Train & Alignment & Edit \\
\rowcolor{gray!10}
Relative Time  & 100\% & 18.1\% & 41.6\%  & 9.5\% \\
\bottomrule
\end{tabular}
}
\label{tab:update}
\vspace{-0.2cm}
\end{table}

\subsubsection{Updating Time Analysis}

A key motivation for using model editing to mitigate cold-start collapse is its low per-item update latency. To highlight this advantage, we conduct an experiment that explicitly compares the total model update time of \ourname, retraining, and finetuning. Specifically, we measure the wall-clock time from completing new-knowledge construction to obtaining the updated model. Although SpecGR does not train on the constructed knowledge and instead relies on an additional alignment task, we still include its training time as the model update time for completeness. For ease of comparison, we normalize all results by the retraining time and report the relative time costs of the other methods. As shown in Table~\ref{tab:update}, \ourname is highly efficient in terms of model update time: as a training-free approach, it incurs only $9.5\%$ of the retraining cost.


\begin{figure}[t]
\centering
\includegraphics[width=1.0\columnwidth]{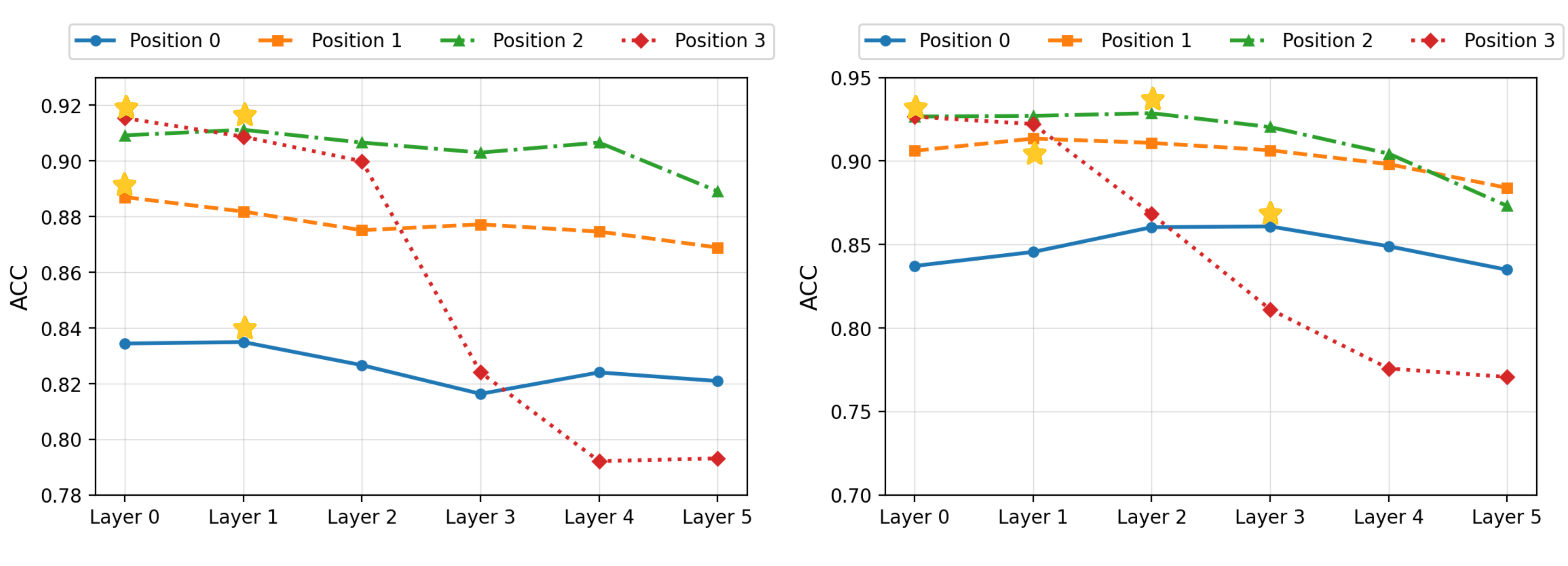}
\vspace{-5px}
\caption{
Classfier Accuracy across layers. The \textbf{left} figure shows the result on Video, while the \textbf{right} figure reports the result on Phone.
}
\label{fig:classifier}
\vspace{-0.6cm}
\end{figure}

\subsubsection{Classifier Accuracy across Layers }


A key component of our method is locating an edit layer for each position. As described in Section~\ref{sec:locatethenedit}, we select the edit laye based on the classification accuracy of a probing classifier. To assess layer-wise sensitivity to new knowledge (cold-start items) and original knowledge (warm items), we report the probing accuracy at each layer.  As shown in Figure~\ref{fig:classifier}, we make two key observations.  \textbf{(1) }Across positions, the classifier exhibits a broadly consistent layer preference: intermediate-to-early layers tend to yield higher probing accuracy. This suggests that the edit, formulated as an equivalent linear transformation, is more effective at separating new versus prior knowledge in these layers, consistent with findings in NLP \citep{meng2022mass}. \textbf{(2)} For position 3, the probing accuracy drops sharply in the later layers. We attribute this to an intrinsic property of our data: at position 3, both new and prior knowledge contain many identical non-semantic tokens, which reduces discriminability in deeper layers.

\section{Conclusion}
Generative recommendation (GR) is a promising paradigm for sequential recommendation, yet we identify a critical bottleneck: \emph{cold-start collapse}, where accuracy on newly introduced items can drop to near zero. Our analysis suggests that GR models often generate the first semantic-ID token correctly but tend to complete sequences with \emph{seen} semantic-ID patterns, making multi-token generation for cold-start items unreliable. To enable timely updates without costly retraining, we propose \ourname, a model-editing framework tailored to GR that treats cold-start semantic-ID patterns as editable knowledge. \ourname performs position-wise, context-to-next-token edits, iteratively injects token-level knowledge to handle the lack of stable token bundles, and adopts a One-One triggering mechanism to avoid cross-position interference during decoding. Experiments on three Amazon categories show that \ourname effectively mitigates cold-start collapse, substantially improving cold-start recommendation while preserving warm-item performance with low overhead.

\section*{Appendix A: Proof Sketch of Theorem \ref{thm:core}}
\label{sec:proof}
Let $z_p^{\text{actual}} = z_p^* + \sum_{q \neq p} \Delta W_q \cdot k_p$ be the actual FFN output. By Assumption \ref{assump:1} and the triangle inequality, the interference from the other 3 positions is bounded by $\| \sum_{q \neq p} \Delta W_q \cdot k_p \|_2 \leq 3\varepsilon$.

Using the Cauchy-Schwarz inequality, the inner products with the target one-hot vector $e_{o_p}$ and any non-target vector $e_o$ are bounded by:
$$
\langle z_p^{\text{actual}}, e_{o_p} \rangle \geq \langle z_p^*, e_{o_p} \rangle - 3\varepsilon, \quad \max_{o \neq o_p} \langle z_p^{\text{actual}}, e_o \rangle \leq \beta + 3\varepsilon,
$$
where $\beta = \max_{o \neq o_p} \langle z_p^*, e_o \rangle$. Given the theorem condition $\varepsilon < \frac{1}{6}(\langle z_p^*, e_{o_p} \rangle - \beta)$, we have $\langle z_p^{\text{actual}}, e_{o_p} \rangle - \max_{o \neq o_p} \langle z_p^{\text{actual}}, e_o \rangle > 0$. Thus, $o_p$ strictly remains the most probable token.

To derive the explicit probability error bound, we define $A = \exp(\langle z_p^*, e_{o_p} \rangle)$ and $B = \sum_{o \neq o_p} \exp(\langle z_p^*, e_o \rangle)$, with $A'$ and $B'$ as their actual counterparts. The inequalities on inner products imply $Ae^{-3\varepsilon} \leq A' \leq Ae^{3\varepsilon}$ and $Be^{-3\varepsilon} \leq B' \leq Be^{3\varepsilon}$. 

The absolute probability difference is:
$$
|p_{\text{actual}} - p_{\text{ideal}}| = \frac{|A'B - AB'|}{(A'+B')(A+B)}.
$$
Maximizing the numerator yields $|A'B - AB'| \leq AB(e^{3\varepsilon} - e^{-3\varepsilon})$, and minimizing the denominator yields $(A'+B')(A+B) \geq e^{-3\varepsilon}(A+B)^2$. Combining these gives:
$$
|p_{\text{actual}} - p_{\text{ideal}}| \leq \frac{AB}{(A+B)^2} \left(e^{6\varepsilon}-1 \right).
$$
By the AM-GM inequality, $\frac{AB}{(A+B)^2} \leq \frac{1}{4}$. Thus, the error is bounded by $\delta(\varepsilon) = \frac{1}{4}(e^{6\varepsilon}-1)$. Applying the first-order Taylor expansion for small $\varepsilon$ yields $\delta(\varepsilon) \approx 1.5\varepsilon$. \hfill $\blacksquare$

\begin{acks}
This work was partially supported by the National Natural Science Foundation of China (No. 62376275, 62472426). Work partially done at Beijing Key Laboratory of Research on Large Models and Intelligent Governance, and Engineering Research Center of Next-Generation Intelligent Search and Recommendation, MOE. Supported by fund for building world-class universities (disciplines) of Renmin University of China.
\end{acks}

\bibliographystyle{ACM-Reference-Format}
\bibliography{ref}


\begin{thebibliography}{43}


\ifx \showCODEN    \undefined \def \showCODEN     #1{\unskip}     \fi
\ifx \showISBNx    \undefined \def \showISBNx     #1{\unskip}     \fi
\ifx \showISBNxiii \undefined \def \showISBNxiii  #1{\unskip}     \fi
\ifx \showISSN     \undefined \def \showISSN      #1{\unskip}     \fi
\ifx \showLCCN     \undefined \def \showLCCN      #1{\unskip}     \fi
\ifx \shownote     \undefined \def \shownote      #1{#1}          \fi
\ifx \showarticletitle \undefined \def \showarticletitle #1{#1}   \fi
\ifx \showURL      \undefined \def \showURL       {\relax}        \fi
\providecommand\bibfield[2]{#2}
\providecommand\bibinfo[2]{#2}
\providecommand\natexlab[1]{#1}
\providecommand\showeprint[2][]{arXiv:#2}

\bibitem[Belinkov(2022)]%
        {belinkov2022probing}
\bibfield{author}{\bibinfo{person}{Yonatan Belinkov}.} \bibinfo{year}{2022}\natexlab{}.
\newblock \showarticletitle{Probing classifiers: Promises, shortcomings, and advances}.
\newblock \bibinfo{journal}{\emph{Computational Linguistics}} \bibinfo{volume}{48}, \bibinfo{number}{1} (\bibinfo{year}{2022}), \bibinfo{pages}{207--219}.
\newblock


\bibitem[Chen et~al\mbox{.}(2023)]%
        {chen2023controllable}
\bibfield{author}{\bibinfo{person}{Sirui Chen}, \bibinfo{person}{Yuan Wang}, \bibinfo{person}{Zijing Wen}, \bibinfo{person}{Zhiyu Li}, \bibinfo{person}{Changshuo Zhang}, \bibinfo{person}{Xiao Zhang}, \bibinfo{person}{Quan Lin}, \bibinfo{person}{Cheng Zhu}, {and} \bibinfo{person}{Jun Xu}.} \bibinfo{year}{2023}\natexlab{}.
\newblock \showarticletitle{Controllable multi-objective re-ranking with policy hypernetworks}. In \bibinfo{booktitle}{\emph{Proceedings of the 29th ACM SIGKDD conference on knowledge discovery and data mining}}. \bibinfo{pages}{3855--3864}.
\newblock


\bibitem[Dai et~al\mbox{.}(2022)]%
        {dai2022knowledge}
\bibfield{author}{\bibinfo{person}{Damai Dai}, \bibinfo{person}{Li Dong}, \bibinfo{person}{Yaru Hao}, \bibinfo{person}{Zhifang Sui}, \bibinfo{person}{Baobao Chang}, {and} \bibinfo{person}{Furu Wei}.} \bibinfo{year}{2022}\natexlab{}.
\newblock \showarticletitle{Knowledge neurons in pretrained transformers}. In \bibinfo{booktitle}{\emph{Proceedings of the 60th Annual Meeting of the Association for Computational Linguistics (Volume 1: Long Papers)}}. \bibinfo{pages}{8493--8502}.
\newblock


\bibitem[Deng et~al\mbox{.}(2025)]%
        {deng2025onerec}
\bibfield{author}{\bibinfo{person}{Jiaxin Deng}, \bibinfo{person}{Shiyao Wang}, \bibinfo{person}{Kuo Cai}, \bibinfo{person}{Lejian Ren}, \bibinfo{person}{Qigen Hu}, \bibinfo{person}{Weifeng Ding}, \bibinfo{person}{Qiang Luo}, {and} \bibinfo{person}{Guorui Zhou}.} \bibinfo{year}{2025}\natexlab{}.
\newblock \showarticletitle{Onerec: Unifying retrieve and rank with generative recommender and iterative preference alignment}.
\newblock \bibinfo{journal}{\emph{arXiv preprint arXiv:2502.18965}} (\bibinfo{year}{2025}).
\newblock


\bibitem[Ding et~al\mbox{.}(2026)]%
        {ding2026inductive}
\bibfield{author}{\bibinfo{person}{Yijie Ding}, \bibinfo{person}{Jiacheng Li}, \bibinfo{person}{Julian McAuley}, {and} \bibinfo{person}{Yupeng Hou}.} \bibinfo{year}{2026}\natexlab{}.
\newblock \showarticletitle{Inductive generative recommendation via retrieval-based speculation}. In \bibinfo{booktitle}{\emph{Proceedings of the AAAI Conference on Artificial Intelligence}}, Vol.~\bibinfo{volume}{40}. \bibinfo{pages}{14675--14683}.
\newblock


\bibitem[Fang et~al\mbox{.}(2024)]%
        {fang2024alphaedit}
\bibfield{author}{\bibinfo{person}{Junfeng Fang}, \bibinfo{person}{Houcheng Jiang}, \bibinfo{person}{Kun Wang}, \bibinfo{person}{Yunshan Ma}, \bibinfo{person}{Shi Jie}, \bibinfo{person}{Xiang Wang}, \bibinfo{person}{Xiangnan He}, {and} \bibinfo{person}{Tat-Seng Chua}.} \bibinfo{year}{2024}\natexlab{}.
\newblock \showarticletitle{Alphaedit: Null-space constrained knowledge editing for language models}.
\newblock \bibinfo{journal}{\emph{arXiv preprint arXiv:2410.02355}} (\bibinfo{year}{2024}).
\newblock


\bibitem[Geva et~al\mbox{.}(2021)]%
        {geva2021transformer}
\bibfield{author}{\bibinfo{person}{Mor Geva}, \bibinfo{person}{Roei Schuster}, \bibinfo{person}{Jonathan Berant}, {and} \bibinfo{person}{Omer Levy}.} \bibinfo{year}{2021}\natexlab{}.
\newblock \showarticletitle{Transformer feed-forward layers are key-value memories}. In \bibinfo{booktitle}{\emph{Proceedings of the 2021 Conference on Empirical Methods in Natural Language Processing}}. \bibinfo{pages}{5484--5495}.
\newblock


\bibitem[Hou et~al\mbox{.}(2023)]%
        {VQRec}
\bibfield{author}{\bibinfo{person}{Yupeng Hou}, \bibinfo{person}{Zhankui He}, \bibinfo{person}{Julian McAuley}, {and} \bibinfo{person}{Wayne~Xin Zhao}.} \bibinfo{year}{2023}\natexlab{}.
\newblock \showarticletitle{Learning vector-quantized item representation for transferable sequential recommenders}. In \bibinfo{booktitle}{\emph{Proceedings of the ACM Web Conference 2023}}. \bibinfo{pages}{1162--1171}.
\newblock


\bibitem[Hou et~al\mbox{.}(2024)]%
        {hou2024bridging}
\bibfield{author}{\bibinfo{person}{Yupeng Hou}, \bibinfo{person}{Jiacheng Li}, \bibinfo{person}{Zhankui He}, \bibinfo{person}{An Yan}, \bibinfo{person}{Xiusi Chen}, {and} \bibinfo{person}{Julian McAuley}.} \bibinfo{year}{2024}\natexlab{}.
\newblock \showarticletitle{Bridging language and items for retrieval and recommendation}.
\newblock \bibinfo{journal}{\emph{arXiv preprint arXiv:2403.03952}} (\bibinfo{year}{2024}).
\newblock


\bibitem[Hou et~al\mbox{.}(2025a)]%
        {RPG}
\bibfield{author}{\bibinfo{person}{Yupeng Hou}, \bibinfo{person}{Jiacheng Li}, \bibinfo{person}{Ashley Shin}, \bibinfo{person}{Jinsung Jeon}, \bibinfo{person}{Abhishek Santhanam}, \bibinfo{person}{Wei Shao}, \bibinfo{person}{Kaveh Hassani}, \bibinfo{person}{Ning Yao}, {and} \bibinfo{person}{Julian McAuley}.} \bibinfo{year}{2025}\natexlab{a}.
\newblock \showarticletitle{Generating long semantic IDs in parallel for recommendation}. In \bibinfo{booktitle}{\emph{Proceedings of the 31st ACM SIGKDD Conference on Knowledge Discovery and Data Mining V. 2}}. \bibinfo{pages}{956--966}.
\newblock


\bibitem[Hou et~al\mbox{.}(2025b)]%
        {hou2025towards}
\bibfield{author}{\bibinfo{person}{Yupeng Hou}, \bibinfo{person}{An Zhang}, \bibinfo{person}{Leheng Sheng}, \bibinfo{person}{Jiancan Wu}, \bibinfo{person}{Xiang Wang}, \bibinfo{person}{Tat-Seng Chua}, {and} \bibinfo{person}{Julian McAuley}.} \bibinfo{year}{2025}\natexlab{b}.
\newblock \showarticletitle{Towards Large Generative Recommendation: A Tokenization Perspective}. In \bibinfo{booktitle}{\emph{Proceedings of the 34th ACM International Conference on Information and Knowledge Management}}. \bibinfo{pages}{6821--6824}.
\newblock


\bibitem[Hua et~al\mbox{.}(2023)]%
        {hua2023index}
\bibfield{author}{\bibinfo{person}{Wenyue Hua}, \bibinfo{person}{Shuyuan Xu}, \bibinfo{person}{Yingqiang Ge}, {and} \bibinfo{person}{Yongfeng Zhang}.} \bibinfo{year}{2023}\natexlab{}.
\newblock \showarticletitle{How to index item ids for recommendation foundation models}. In \bibinfo{booktitle}{\emph{Proceedings of the Annual International ACM SIGIR Conference on Research and Development in Information Retrieval in the Asia Pacific Region}}. \bibinfo{pages}{195--204}.
\newblock


\bibitem[Jiang et~al\mbox{.}(2025)]%
        {jiang2025anyedit}
\bibfield{author}{\bibinfo{person}{Houcheng Jiang}, \bibinfo{person}{Junfeng Fang}, \bibinfo{person}{Ningyu Zhang}, \bibinfo{person}{Guojun Ma}, \bibinfo{person}{Mingyang Wan}, \bibinfo{person}{Xiang Wang}, \bibinfo{person}{Xiangnan He}, {and} \bibinfo{person}{Tat-Seng Chua}.} \bibinfo{year}{2025}\natexlab{}.
\newblock \showarticletitle{AnyEdit: Edit Any Knowledge Encoded in Language Models}.
\newblock \bibinfo{journal}{\emph{CoRR}} (\bibinfo{year}{2025}).
\newblock


\bibitem[Kang and McAuley(2018)]%
        {SASREC}
\bibfield{author}{\bibinfo{person}{Wang-Cheng Kang} {and} \bibinfo{person}{Julian McAuley}.} \bibinfo{year}{2018}\natexlab{}.
\newblock \showarticletitle{Self-attentive sequential recommendation}. In \bibinfo{booktitle}{\emph{2018 IEEE International Conference on Data Mining (ICDM)}}. IEEE, \bibinfo{pages}{197--206}.
\newblock


\bibitem[Li et~al\mbox{.}(2023)]%
        {li2023inference}
\bibfield{author}{\bibinfo{person}{Kenneth Li}, \bibinfo{person}{Oam Patel}, \bibinfo{person}{Fernanda Vi{\'e}gas}, \bibinfo{person}{Hanspeter Pfister}, {and} \bibinfo{person}{Martin Wattenberg}.} \bibinfo{year}{2023}\natexlab{}.
\newblock \showarticletitle{Inference-time intervention: Eliciting truthful answers from a language model}.
\newblock \bibinfo{journal}{\emph{Advances in Neural Information Processing Systems}}  \bibinfo{volume}{36} (\bibinfo{year}{2023}).
\newblock


\bibitem[Loshchilov and Hutter(2017)]%
        {loshchilov2017decoupled}
\bibfield{author}{\bibinfo{person}{Ilya Loshchilov} {and} \bibinfo{person}{Frank Hutter}.} \bibinfo{year}{2017}\natexlab{}.
\newblock \showarticletitle{Decoupled weight decay regularization}.
\newblock \bibinfo{journal}{\emph{arXiv preprint arXiv:1711.05101}} (\bibinfo{year}{2017}).
\newblock


\bibitem[Meng et~al\mbox{.}(2022a)]%
        {meng2022locating}
\bibfield{author}{\bibinfo{person}{Kevin Meng}, \bibinfo{person}{David Bau}, \bibinfo{person}{Alex Andonian}, {and} \bibinfo{person}{Yonatan Belinkov}.} \bibinfo{year}{2022}\natexlab{a}.
\newblock \showarticletitle{Locating and editing factual associations in gpt}.
\newblock \bibinfo{journal}{\emph{Advances in neural information processing systems}}  \bibinfo{volume}{35} (\bibinfo{year}{2022}), \bibinfo{pages}{17359--17372}.
\newblock


\bibitem[Meng et~al\mbox{.}(2022b)]%
        {meng2022mass}
\bibfield{author}{\bibinfo{person}{Kevin Meng}, \bibinfo{person}{Arnab~Sen Sharma}, \bibinfo{person}{Alex Andonian}, \bibinfo{person}{Yonatan Belinkov}, {and} \bibinfo{person}{David Bau}.} \bibinfo{year}{2022}\natexlab{b}.
\newblock \showarticletitle{Mass-editing memory in a transformer}.
\newblock \bibinfo{journal}{\emph{arXiv preprint arXiv:2210.07229}} (\bibinfo{year}{2022}).
\newblock


\bibitem[Ni et~al\mbox{.}(2021)]%
        {ni2021sentence}
\bibfield{author}{\bibinfo{person}{Jianmo Ni}, \bibinfo{person}{Gustavo~Hernandez Abrego}, \bibinfo{person}{Noah Constant}, \bibinfo{person}{Ji Ma}, \bibinfo{person}{Keith~B Hall}, \bibinfo{person}{Daniel Cer}, {and} \bibinfo{person}{Yinfei Yang}.} \bibinfo{year}{2021}\natexlab{}.
\newblock \showarticletitle{Sentence-t5: Scalable sentence encoders from pre-trained text-to-text models}.
\newblock \bibinfo{journal}{\emph{arXiv preprint arXiv:2108.08877}} (\bibinfo{year}{2021}).
\newblock


\bibitem[Qin et~al\mbox{.}(2025a)]%
        {qin2025maps}
\bibfield{author}{\bibinfo{person}{Weicong Qin}, \bibinfo{person}{Yi Xu}, \bibinfo{person}{Weijie Yu}, \bibinfo{person}{Chenglei Shen}, \bibinfo{person}{Ming He}, \bibinfo{person}{Jianping Fan}, \bibinfo{person}{Xiao Zhang}, {and} \bibinfo{person}{Jun Xu}.} \bibinfo{year}{2025}\natexlab{a}.
\newblock \showarticletitle{Maps: Motivation-aware personalized search via llm-driven consultation alignment}. In \bibinfo{booktitle}{\emph{Proceedings of the 63rd Annual Meeting of the Association for Computational Linguistics (Volume 1: Long Papers)}}. \bibinfo{pages}{3039--3051}.
\newblock


\bibitem[Qin et~al\mbox{.}(2025b)]%
        {qin2025more}
\bibfield{author}{\bibinfo{person}{Weicong Qin}, \bibinfo{person}{Yi Xu}, \bibinfo{person}{Weijie Yu}, \bibinfo{person}{Chenglei Shen}, \bibinfo{person}{Xiao Zhang}, \bibinfo{person}{Ming He}, \bibinfo{person}{Jianping Fan}, {and} \bibinfo{person}{Jun Xu}.} \bibinfo{year}{2025}\natexlab{b}.
\newblock \showarticletitle{More: A mixture of reflectors framework for large language model-based sequential recommendation}. In \bibinfo{booktitle}{\emph{Proceedings of the Nineteenth ACM Conference on Recommender Systems}}. \bibinfo{pages}{299--308}.
\newblock


\bibitem[Qu et~al\mbox{.}(2025)]%
        {qu2025bridging}
\bibfield{author}{\bibinfo{person}{Changle Qu}, \bibinfo{person}{Liqin Zhao}, \bibinfo{person}{Yanan Niu}, \bibinfo{person}{Xiao Zhang}, {and} \bibinfo{person}{Jun Xu}.} \bibinfo{year}{2025}\natexlab{}.
\newblock \showarticletitle{Bridging Short Videos and Streamers with Multi-Graph Contrastive Learning for Live Streaming Recommendation}. In \bibinfo{booktitle}{\emph{Proceedings of the 48th International ACM SIGIR Conference on Research and Development in Information Retrieval}}. \bibinfo{pages}{2059--2069}.
\newblock


\bibitem[Rajput et~al\mbox{.}(2023)]%
        {TIGER}
\bibfield{author}{\bibinfo{person}{Shashank Rajput}, \bibinfo{person}{Nikhil Mehta}, \bibinfo{person}{Anima Singh}, \bibinfo{person}{Raghunandan Hulikal~Keshavan}, \bibinfo{person}{Trung Vu}, \bibinfo{person}{Lukasz Heldt}, \bibinfo{person}{Lichan Hong}, \bibinfo{person}{Yi Tay}, \bibinfo{person}{Vinh Tran}, \bibinfo{person}{Jonah Samost}, {et~al\mbox{.}}} \bibinfo{year}{2023}\natexlab{}.
\newblock \showarticletitle{Recommender systems with generative retrieval}.
\newblock \bibinfo{journal}{\emph{Advances in Neural Information Processing Systems}}  \bibinfo{volume}{36} (\bibinfo{year}{2023}), \bibinfo{pages}{10299--10315}.
\newblock


\bibitem[Shen et~al\mbox{.}(2026a)]%
        {shen2026enhancing}
\bibfield{author}{\bibinfo{person}{Chenglei Shen}, \bibinfo{person}{Yi Zhan}, \bibinfo{person}{Weijie Yu}, \bibinfo{person}{Xiao Zhang}, {and} \bibinfo{person}{Jun Xu}.} \bibinfo{year}{2026}\natexlab{a}.
\newblock \showarticletitle{Enhancing Bandit Algorithms with LLMs for Time-varying User Preferences in Streaming Recommendations}.
\newblock \bibinfo{journal}{\emph{ACM Transactions on Information Systems}} \bibinfo{volume}{44}, \bibinfo{number}{3} (\bibinfo{year}{2026}), \bibinfo{pages}{1--30}.
\newblock


\bibitem[Shen et~al\mbox{.}(2026b)]%
        {shen2026survey}
\bibfield{author}{\bibinfo{person}{Chenglei Shen}, \bibinfo{person}{Xiao Zhang}, \bibinfo{person}{Teng Shi}, \bibinfo{person}{Changshuo Zhang}, \bibinfo{person}{Guofu Xie}, \bibinfo{person}{Jun Xu}, \bibinfo{person}{Ming He}, {and} \bibinfo{person}{Jianping Fan}.} \bibinfo{year}{2026}\natexlab{b}.
\newblock \showarticletitle{A survey of controllable learning: Methods and applications in information retrieval}.
\newblock \bibinfo{journal}{\emph{Frontiers of Computer Science}} \bibinfo{volume}{20}, \bibinfo{number}{10} (\bibinfo{year}{2026}), \bibinfo{pages}{2010619}.
\newblock


\bibitem[Shen et~al\mbox{.}(2023)]%
        {shen2023hyperbandit}
\bibfield{author}{\bibinfo{person}{Chenglei Shen}, \bibinfo{person}{Xiao Zhang}, \bibinfo{person}{Wei Wei}, {and} \bibinfo{person}{Jun Xu}.} \bibinfo{year}{2023}\natexlab{}.
\newblock \showarticletitle{Hyperbandit: Contextual bandit with hypernewtork for time-varying user preferences in streaming recommendation}. In \bibinfo{booktitle}{\emph{Proceedings of the 32nd ACM International Conference on Information and Knowledge Management}}. \bibinfo{pages}{2239--2248}.
\newblock


\bibitem[Shen et~al\mbox{.}(2025)]%
        {shen2025paragon}
\bibfield{author}{\bibinfo{person}{Chenglei Shen}, \bibinfo{person}{Jiahao Zhao}, \bibinfo{person}{Xiao Zhang}, \bibinfo{person}{Weijie Yu}, \bibinfo{person}{Ming He}, {and} \bibinfo{person}{Jianping Fan}.} \bibinfo{year}{2025}\natexlab{}.
\newblock \showarticletitle{Paragon: Parameter Generation for Controllable Multi-Task Recommendation}. In \bibinfo{booktitle}{\emph{Proceedings of the Nineteenth ACM Conference on Recommender Systems}}. \bibinfo{pages}{370--380}.
\newblock


\bibitem[Shi et~al\mbox{.}(2025)]%
        {shi2025llada}
\bibfield{author}{\bibinfo{person}{Teng Shi}, \bibinfo{person}{Chenglei Shen}, \bibinfo{person}{Weijie Yu}, \bibinfo{person}{Shen Nie}, \bibinfo{person}{Chongxuan Li}, \bibinfo{person}{Xiao Zhang}, \bibinfo{person}{Ming He}, \bibinfo{person}{Yan Han}, {and} \bibinfo{person}{Jun Xu}.} \bibinfo{year}{2025}\natexlab{}.
\newblock \showarticletitle{LLaDA-Rec: Discrete Diffusion for Parallel Semantic ID Generation in Generative Recommendation}.
\newblock \bibinfo{journal}{\emph{arXiv preprint arXiv:2511.06254}} (\bibinfo{year}{2025}).
\newblock


\bibitem[Si et~al\mbox{.}(2024)]%
        {si2024generative}
\bibfield{author}{\bibinfo{person}{Zihua Si}, \bibinfo{person}{Zhongxiang Sun}, \bibinfo{person}{Jiale Chen}, \bibinfo{person}{Guozhang Chen}, \bibinfo{person}{Xiaoxue Zang}, \bibinfo{person}{Kai Zheng}, \bibinfo{person}{Yang Song}, \bibinfo{person}{Xiao Zhang}, \bibinfo{person}{Jun Xu}, {and} \bibinfo{person}{Kun Gai}.} \bibinfo{year}{2024}\natexlab{}.
\newblock \showarticletitle{Generative retrieval with semantic tree-structured identifiers and contrastive learning}. In \bibinfo{booktitle}{\emph{Proceedings of the 2024 Annual International ACM SIGIR Conference on Research and Development in Information Retrieval in the Asia Pacific Region}}. \bibinfo{pages}{154--163}.
\newblock


\bibitem[Sun et~al\mbox{.}(2019a)]%
        {sun2019bert4rec}
\bibfield{author}{\bibinfo{person}{Fei Sun}, \bibinfo{person}{Jun Liu}, \bibinfo{person}{Jian Wu}, \bibinfo{person}{Changhua Pei}, \bibinfo{person}{Xiao Lin}, \bibinfo{person}{Wenwu Ou}, {and} \bibinfo{person}{Peng Jiang}.} \bibinfo{year}{2019}\natexlab{a}.
\newblock \showarticletitle{BERT4Rec: Sequential recommendation with bidirectional encoder representations from transformer}. In \bibinfo{booktitle}{\emph{Proceedings of the 28th ACM international conference on information and knowledge management}}. \bibinfo{pages}{1441--1450}.
\newblock


\bibitem[Sun et~al\mbox{.}(2019b)]%
        {BERT4REC}
\bibfield{author}{\bibinfo{person}{Fei Sun}, \bibinfo{person}{Jun Liu}, \bibinfo{person}{Jian Wu}, \bibinfo{person}{Changhua Pei}, \bibinfo{person}{Xiao Lin}, \bibinfo{person}{Wenwu Ou}, {and} \bibinfo{person}{Peng Jiang}.} \bibinfo{year}{2019}\natexlab{b}.
\newblock \showarticletitle{BERT4Rec: Sequential Recommendation with Bidirectional Encoder Representations from Transformer}. In \bibinfo{booktitle}{\emph{Proceedings of the 28th ACM International Conference on Information and Knowledge Management}} (Beijing, China) \emph{(\bibinfo{series}{CIKM '19})}. \bibinfo{publisher}{ACM}, \bibinfo{address}{New York, NY, USA}, \bibinfo{pages}{1441--1450}.
\newblock
\showISBNx{978-1-4503-6976-3}


\bibitem[Wang et~al\mbox{.}(2024a)]%
        {LETTER}
\bibfield{author}{\bibinfo{person}{Wenjie Wang}, \bibinfo{person}{Honghui Bao}, \bibinfo{person}{Xinyu Lin}, \bibinfo{person}{Jizhi Zhang}, \bibinfo{person}{Yongqi Li}, \bibinfo{person}{Fuli Feng}, \bibinfo{person}{See-Kiong Ng}, {and} \bibinfo{person}{Tat-Seng Chua}.} \bibinfo{year}{2024}\natexlab{a}.
\newblock \showarticletitle{Learnable item tokenization for generative recommendation}. In \bibinfo{booktitle}{\emph{Proceedings of the 33rd ACM International Conference on Information and Knowledge Management}}. \bibinfo{pages}{2400--2409}.
\newblock


\bibitem[Wang et~al\mbox{.}(2024b)]%
        {wang2024eager}
\bibfield{author}{\bibinfo{person}{Ye Wang}, \bibinfo{person}{Jiahao Xun}, \bibinfo{person}{Minjie Hong}, \bibinfo{person}{Jieming Zhu}, \bibinfo{person}{Tao Jin}, \bibinfo{person}{Wang Lin}, \bibinfo{person}{Haoyuan Li}, \bibinfo{person}{Linjun Li}, \bibinfo{person}{Yan Xia}, \bibinfo{person}{Zhou Zhao}, {et~al\mbox{.}}} \bibinfo{year}{2024}\natexlab{b}.
\newblock \showarticletitle{Eager: Two-stream generative recommender with behavior-semantic collaboration}. In \bibinfo{booktitle}{\emph{Proceedings of the 30th ACM SIGKDD Conference on Knowledge Discovery and Data Mining}}. \bibinfo{pages}{3245--3254}.
\newblock


\bibitem[Yang et~al\mbox{.}(2024a)]%
        {yang2024unifying}
\bibfield{author}{\bibinfo{person}{Liu Yang}, \bibinfo{person}{Fabian Paischer}, \bibinfo{person}{Kaveh Hassani}, \bibinfo{person}{Jiacheng Li}, \bibinfo{person}{Shuai Shao}, \bibinfo{person}{Zhang~Gabriel Li}, \bibinfo{person}{Yun He}, \bibinfo{person}{Xue Feng}, \bibinfo{person}{Nima Noorshams}, \bibinfo{person}{Sem Park}, {et~al\mbox{.}}} \bibinfo{year}{2024}\natexlab{a}.
\newblock \showarticletitle{Unifying generative and dense retrieval for sequential recommendation}.
\newblock \bibinfo{journal}{\emph{arXiv preprint arXiv:2411.18814}} (\bibinfo{year}{2024}).
\newblock


\bibitem[Yang et~al\mbox{.}(2024b)]%
        {yang2024ttt4rec}
\bibfield{author}{\bibinfo{person}{Zhaoqi Yang}, \bibinfo{person}{Yanan Wang}, {and} \bibinfo{person}{Yong Ge}.} \bibinfo{year}{2024}\natexlab{b}.
\newblock \showarticletitle{TTT4Rec: A Test-Time Training Approach for Rapid Adaption in Sequential Recommendation}.
\newblock \bibinfo{journal}{\emph{arXiv preprint arXiv:2409.19142}} (\bibinfo{year}{2024}).
\newblock


\bibitem[Zeng et~al\mbox{.}(2026)]%
        {zeng2026raie}
\bibfield{author}{\bibinfo{person}{Jin Zeng}, \bibinfo{person}{Yupeng Qi}, \bibinfo{person}{Hui Li}, \bibinfo{person}{Chengming Li}, \bibinfo{person}{Ziyu Lyu}, \bibinfo{person}{Lixin Cui}, {and} \bibinfo{person}{Lu Bai}.} \bibinfo{year}{2026}\natexlab{}.
\newblock \showarticletitle{RAIE: Region-Aware Incremental Preference Editing with LoRA for LLM-based Recommendation}.
\newblock \bibinfo{journal}{\emph{arXiv preprint arXiv:2603.00638}} (\bibinfo{year}{2026}).
\newblock


\bibitem[Zhai et~al\mbox{.}(2024)]%
        {zhai2024actions}
\bibfield{author}{\bibinfo{person}{Jiaqi Zhai}, \bibinfo{person}{Lucy Liao}, \bibinfo{person}{Xing Liu}, \bibinfo{person}{Yueming Wang}, \bibinfo{person}{Rui Li}, \bibinfo{person}{Xuan Cao}, \bibinfo{person}{Leon Gao}, \bibinfo{person}{Zhaojie Gong}, \bibinfo{person}{Fangda Gu}, \bibinfo{person}{Jiayuan He}, {et~al\mbox{.}}} \bibinfo{year}{2024}\natexlab{}.
\newblock \showarticletitle{Actions Speak Louder than Words: Trillion-Parameter Sequential Transducers for Generative Recommendations}. In \bibinfo{booktitle}{\emph{International Conference on Machine Learning}}. PMLR, \bibinfo{pages}{58484--58509}.
\newblock


\bibitem[Zhang et~al\mbox{.}(2025)]%
        {zhang2025test}
\bibfield{author}{\bibinfo{person}{Changshuo Zhang}, \bibinfo{person}{Xiao Zhang}, \bibinfo{person}{Teng Shi}, \bibinfo{person}{Jun Xu}, {and} \bibinfo{person}{Ji-Rong Wen}.} \bibinfo{year}{2025}\natexlab{}.
\newblock \showarticletitle{Test-Time Alignment with State Space Model for Tracking User Interest Shifts in Sequential Recommendation}. In \bibinfo{booktitle}{\emph{Proceedings of the Nineteenth ACM Conference on Recommender Systems}}. \bibinfo{pages}{461--471}.
\newblock


\bibitem[Zhang et~al\mbox{.}(2026)]%
        {zhang2026model}
\bibfield{author}{\bibinfo{person}{Zhen Zhang}, \bibinfo{person}{Zihan Wang}, \bibinfo{person}{Xinyu Ma}, \bibinfo{person}{Shuaiqiang Wang}, \bibinfo{person}{Dawei Yin}, \bibinfo{person}{Xin Xin}, \bibinfo{person}{Pengjie Ren}, \bibinfo{person}{Maarten de Rijke}, {and} \bibinfo{person}{Zhaochun Ren}.} \bibinfo{year}{2026}\natexlab{}.
\newblock \showarticletitle{Model Editing for New Document Integration in Generative Information Retrieval}.
\newblock \bibinfo{journal}{\emph{arXiv preprint arXiv:2603.02773}} (\bibinfo{year}{2026}).
\newblock


\bibitem[Zhao et~al\mbox{.}(2023)]%
        {zhao2023survey}
\bibfield{author}{\bibinfo{person}{Wayne~Xin Zhao}, \bibinfo{person}{Kun Zhou}, \bibinfo{person}{Junyi Li}, \bibinfo{person}{Tianyi Tang}, \bibinfo{person}{Xiaolei Wang}, \bibinfo{person}{Yupeng Hou}, \bibinfo{person}{Yingqian Min}, \bibinfo{person}{Beichen Zhang}, \bibinfo{person}{Junjie Zhang}, \bibinfo{person}{Zican Dong}, {et~al\mbox{.}}} \bibinfo{year}{2023}\natexlab{}.
\newblock \showarticletitle{A survey of large language models}.
\newblock \bibinfo{journal}{\emph{arXiv preprint arXiv:2303.18223}} \bibinfo{volume}{1}, \bibinfo{number}{2} (\bibinfo{year}{2023}).
\newblock


\bibitem[Zheng et~al\mbox{.}(2024)]%
        {LC_Rec}
\bibfield{author}{\bibinfo{person}{Bowen Zheng}, \bibinfo{person}{Yupeng Hou}, \bibinfo{person}{Hongyu Lu}, \bibinfo{person}{Yu Chen}, \bibinfo{person}{Wayne~Xin Zhao}, \bibinfo{person}{Ming Chen}, {and} \bibinfo{person}{Ji-Rong Wen}.} \bibinfo{year}{2024}\natexlab{}.
\newblock \showarticletitle{Adapting large language models by integrating collaborative semantics for recommendation}. In \bibinfo{booktitle}{\emph{2024 IEEE 40th International Conference on Data Engineering (ICDE)}}. IEEE, \bibinfo{pages}{1435--1448}.
\newblock


\bibitem[Zhou et~al\mbox{.}(2025)]%
        {zhou2025onerec}
\bibfield{author}{\bibinfo{person}{Guorui Zhou}, \bibinfo{person}{Jiaxin Deng}, \bibinfo{person}{Jinghao Zhang}, \bibinfo{person}{Kuo Cai}, \bibinfo{person}{Lejian Ren}, \bibinfo{person}{Qiang Luo}, \bibinfo{person}{Qianqian Wang}, \bibinfo{person}{Qigen Hu}, \bibinfo{person}{Rui Huang}, \bibinfo{person}{Shiyao Wang}, {et~al\mbox{.}}} \bibinfo{year}{2025}\natexlab{}.
\newblock \showarticletitle{OneRec Technical Report}.
\newblock \bibinfo{journal}{\emph{arXiv preprint arXiv:2506.13695}} (\bibinfo{year}{2025}).
\newblock


\bibitem[Zhu et~al\mbox{.}(2024)]%
        {zhu2024cost}
\bibfield{author}{\bibinfo{person}{Jieming Zhu}, \bibinfo{person}{Mengqun Jin}, \bibinfo{person}{Qijiong Liu}, \bibinfo{person}{Zexuan Qiu}, \bibinfo{person}{Zhenhua Dong}, {and} \bibinfo{person}{Xiu Li}.} \bibinfo{year}{2024}\natexlab{}.
\newblock \showarticletitle{Cost: Contrastive quantization based semantic tokenization for generative recommendation}. In \bibinfo{booktitle}{\emph{Proceedings of the 18th ACM Conference on Recommender Systems}}. \bibinfo{pages}{969--974}.
\newblock


\end{thebibliography}

\end{document}